\documentclass{emulateapj}

\shorttitle{Stellar Feedback in HII Regions}
\shortauthors{LOPEZ ET AL.}

\newcommand{\ltsima}{$\; \buildrel < \over \sim \;$}
\newcommand{\simlt}{\lower.5ex\hbox{\ltsima}}

\newcommand{\ls}{{_<\atop^{\sim}}}

\newcommand{\gs}{{_>\atop^{\sim}}}

\newcommand{\hii}{H~{\sc ii}}

\def\arcmin{\hbox{$^\prime$}}
\def\arcsec{\hbox{$^{\prime\prime}$}}

\begin{document}

\title{The Role of Stellar Feedback in the Dynamics of HII Regions}

\author{Laura A. Lopez\altaffilmark{1,5,6}, Mark R. Krumholz\altaffilmark{2}, Alberto D. Bolatto\altaffilmark{3}, J. Xavier Prochaska\altaffilmark{2,4}, Enrico Ramirez-Ruiz\altaffilmark{2}, Daniel Castro\altaffilmark{1}}
%%%
\altaffiltext{1}{MIT-Kavli Institute for Astrophysics and Space Research, 77 Massachusetts Avenue, 37-664H, Cambridge MA 02139, USA}
\altaffiltext{2}{Department of Astronomy and Astrophysics, University of California Santa Cruz, 1156 High Street, Santa Cruz, CA 95060, USA}
\altaffiltext{3}{Department of Astronomy, University of Maryland, College Park, MD 20742, USA}
\altaffiltext{4}{UCO/Lick Observatory}
\altaffiltext{5}{NASA Einstein Fellow}
\altaffiltext{6}{Pappalardo Fellow in Physics}

\email{lopez@space.mit.edu}

\begin{abstract}

Stellar feedback is often cited as the biggest uncertainty in galaxy formation models today. This uncertainty stems from a dearth of observational constraints as well as the great dynamic range between the small scales ($\ls$1 pc) where the feedback originates and the large scales of galaxies ($\gs$1 kpc) that are shaped by this feedback. To bridge this divide, in this paper we aim to assess observationally the role of stellar feedback at the intermediate scales of \hii\ regions ($\sim$10--100 pc). In particular, we employ multiwavelength data to examine several stellar feedback mechanisms in a sample of 32 \hii\ regions (with ages $\sim$3--10 Myr) in the Large and Small Magellanic Clouds (LMC and SMC, respectively). Using optical, infrared, radio, and X-ray images, we measure the pressures exerted on the shells from the direct stellar radiation, the dust-processed radiation, the warm ionized gas, and the hot X-ray emitting gas. We find that the warm ionized gas dominates over the other terms in all of the sources, although two have comparable dust-processed radiation pressures to their warm gas pressures. The hot gas pressures are comparatively weak, while the direct radiation pressures are 1--2 orders of magnitude below the other terms. We discuss the implications of these results, particularly highlighting evidence for hot gas leakage from the \hii\ shells and regarding the momentum deposition from the dust-processed radiation to the warm gas. Furthermore, we emphasize that similar observational work should be done on very young \hii\ regions to test whether direct radiation pressure and hot gas can drive the dynamics at early times. 

\end{abstract} 


\section{Introduction}

Stellar feedback -- the injection of energy and momentum by stars -- originates at the small scales of star clusters ($\ls$1 pc), yet it shapes the interstellar medium (ISM) on large scales ($\gs$1 kpc). At large scales, stellar feedback is necessary in order to form realistic galaxies in simulations and to account for observed galaxy properties. In the absence of feedback, baryonic matter cools rapidly and efficiently forms stars, producing an order of magnitude too much stellar mass and consuming most available gas in the galaxy (e.g., \citealt{white78,keres09}). Stellar feedback prevents this ``cooling catastrophe'' by heating gas as well as removing low angular momentum baryons from galactic centers, thereby allowing only a small fraction of the baryonic budget of dark matter halos to be converted to stars. The removal of baryons may also flatten the dark matter mass profile, critical to form bulgeless dwarf galaxies (e.g., \citealt{mash08,governato10,governato12}). Furthermore, stellar feedback possibly drives kpc-scale galactic winds and outflows (see \citealt{veilleux05} for a review) which have been frequently observed in local galaxies (e.g., \citealt{bland88,martin99,strickland04}) as well as in galaxies at moderate to high redshift (e.g., \citealt{ajiki02,frye02,shapley03,rubin13}). 

%Hi, Overheardonastroph! We hope your enjoy our mad latexing skillz.

At the smaller scales of star clusters and giant molecular clouds (GMCs), newborn stars dramatically influence their environments. Observational evidence suggests that only a small fraction ($\sim$1--2\%) of GMC mass is converted to stars per cloud free-fall time (e.g., \citealt{zuckerman74,krumholztan07,evans09,heiderman10,krumholz12}). This inefficiency can be attributed to stellar feedback processes of \hii\ regions that act to disrupt and ultimately to destroy their host clouds (e.g., \citealt{whitworth79,matzner02,dale05,krumholz06,vaz10,goldbaum11,dale12,dale13}). In addition to the pressure of the warm ionized H~\textsc{ii} region gas itself, there are several other forms of stellar feedback that can drive the dynamics of H~\textsc{ii} regions and deposit energy and momentum in the surrounding ISM: the direct radiation of stars (e.g., \citealt{km09,fall10,murray10,hopkins11}), the dust-processed infrared radiation (e.g., \citealt{thompson05,murray10,andrews11}), stellar winds and supernovae (SNe; e.g., \citealt{yorke89,harperclark09,rogers13}), and protostellar outflows/jets (e.g., \citealt{quillen05,cunningham06,li06,nakamura08,wang10}).

From a theoretical perspective, SNe were the first feedback mechanism to be considered as a means to remove gas from low-mass galaxies (e.g., \citealt{dekel86}) and to prevent the cooling catastrophe (e.g., \citealt{white91}). However, resolution limitations precluded the explicit modeling of individual SNe in galaxy formation simulations, so phenomenological prescriptions were employed to account for ``sub-grid'' feedback (e.g., \citealt{katz92,navarro93,mihos94}). Since then, extensive work has been done to improve and to compare these sub-grid models (e.g., \citealt{thacker00,springel03,saitoh08,tey10,hopkins11,sca12,stinson12,aumer13,kim14}). Furthermore, the use of ``zoom-in'' simulations (which can model feedback physics down to $\gs$1 pc scale) has enabled the modeling of several modes of feedback simultaneously (e.g., \citealt{agertz13,hopkins13,renaud13,stinson13,agertz14,ceverino14}).

While simulations are beginning to incorporate many feedback mechanisms, most observational work focuses on the effects of the individual modes. Consequently, the relative contribution of these components and which processes dominate in different conditions remains uncertain. To address this issue, we recently employed multiwavelength imaging of the giant \hii\ region N157 (30~Doradus; ``30 Dor'' hereafter) to assess the dynamical role of several stellar feedback mechanisms in driving the shell expansion \citep{lopez11}. In particular, we measured the pressures associated with the different feedback modes across 441 regions to map the pressure components as a function of position; we considered the direct radiation pressure exerted by the light from massive stars, the dust-processed radiation pressure, the warm ionized ($\sim10^{4}$~K) gas pressure, and the hot shocked ($\sim10^{7}$~K) gas pressure from stellar winds and SNe. We found that the direct radiation pressure from massive stars dominates at distances $\ls$75 pc from the central star cluster R136, while the warm ($\sim10^4$~K) ionized gas pressure dominates at larger radii. By comparison, the dust-processed radiation pressure and the hot ($\sim10^{7}$~K) gas pressure are weak and are not dynamically important on the large scale (although small bubbles of the hot gas can have significant pressures -- \citealt{pellegrini11}; see Appendix~\ref{app:hot gas} of this paper for a discussion on how choice of hot gas filling factor is critical when evaluating the dynamical role of hot gas). 

In this paper, we extend the methodology applied to 30~Dor to a larger sample of 32 \hii\ regions in the Large and Small Magellanic Clouds (LMC and SMC, respectively), with the aim of probing how stellar feedback properties vary between sources. The organization of this paper is as follows. Section~\ref{sec:sample} describes our LMC and SMC \hii\ region sample and the data we have employed for our analyses. Section~\ref{sec:method} outlines the methods we have used to assess the dynamical role of several stellar feedback mechanisms in the 32 sources. Section~\ref{sec:results} presents the results from these analyses, and Section~\ref{sec:discussion} explores implications of our findings related to the importance of radiation pressure (Section~\ref{sec:radpressure}), the confinement of hot gas in the \hii\ regions (Section~\ref{sec:leakage}) and the momentum deposition of the dust-processed radiation to the warm gas (Section~\ref{sec:dusty}). Finally, we summarize this work in Section~\ref{sec:summary}.

~~

\section{Sample \& Data} \label{sec:sample}

For our feedback analyses, we selected the 16 LMC and 16 SMC \hii\ regions of \cite{lawton}, who chose sources based on their bright 24$\mu$m and H$\alpha$ emission and which are distributed throughout these galaxies. We opted to include sources based on both IR and H$\alpha$, since bright H$\alpha$ emission alone is not unique to \hii\ regions. For example, several of the emission nebulae identified by \cite{kenn86} are now known to be supernova remnants. Furthermore, bright 24$\mu$m emission arises from stochastically heated small dust grains (i.e., dust is heated by collisions with starlight photons: e.g., \citealt{draine01}), so it is well-correlated with \hii\ regions within the Milky Way and other galaxies. 

\begin{deluxetable*}{lccccc} 
\tablecolumns{6}
\tablewidth{0pt} \tablecaption{Sample of \hii\ Regions}
\tablehead{\colhead{Source} & \colhead{Alt Name} & \colhead{RA} & \colhead{Dec} & \colhead{Radius\tablenotemark{a}} & \colhead{Radius\tablenotemark{a}} \\
\colhead{} & \colhead{} & \colhead{(J2000)} & \colhead{(J2000)} & \colhead{(arcmin)} & \colhead{(pc)}}
\startdata
\cutinhead{LMC Sources}
N4 & DEM L008 & 04:52:09 & $-$66:55:13 & 0.7 & 10.2 \\ 
N11 & DEM L034, L041 & 04:56:41 & $-$66:27:19 & 10.0 & 145 \\
N30 & DEM L105, L106 & 05:13:51 & $-$67:27:22 & 3.1 & 45.1 \\
N44 & DEM L150 & 05:22:16 & $-$67:57:09 &  7.1 & 103 \\
N48 & DEM L189 & 05:25:50 & $-$66:15:03 & 5.2 & 75.6 \\
N55 & DEM L227, L228 & 05:32:33 & $-$66:27:20 & 3.6 & 52.4 \\
N59 & DEM L241 & 05:35:24 & $-$67:33:22 & 3.9 & 56.7 \\
N79 & DEM L010 & 04:52:04 & $-$69:22:34 & 4.4 & 64.0 \\
N105 & DEM L086 & 05:09:56 & $-$68:54:03 & 2.9 & 42.2 \\
N119 & DEM L132 & 05:18:45 & $-$69:14:03 & 5.9 & 85.8 \\
N144 & DEM L199 & 05:26:38 & $-$68:49:55 & 4.9 & 71.3  \\
N157 & 30 Dor & 05:38:36 & $-$69:05:33 & 6.8 & 98.9 \\ 
N160 & & 05:40:22 & $-$69:37:35 & 5.0 & 40.0 \\
N180 & DEM L322, L323 & 05:48:52 & $-$70:03:51 & 2.7 & 39.3 \\
N191 & DEM L064 & 05:04:35 & $-$70:54:27 & 2.1 & 30.5 \\
N206 & DEM L221 & 05:30:38 & $-$71:03:53 & 7.7 & 112 \\
\cutinhead{SMC Sources}
DEM~S74 & & 00:53:14 & $-$73:12:18 & 2.7 & 47.9 \\
N13 & & 00:45:23 & $-$73:22:38 & 0.5 & 8.87  \\
N17 & & 00:46:41 & $-$73:31:38 & 1.5 & 26.6 \\
N19 & & 00:48:23 & $-$73:05:54 & 0.7 & 12.4 \\
N22 & & 00:48:09 & $-$73:14:56 & 0.9 & 16.0  \\
N36 & & 00:50:26 & $-$72:52:59 & 2.5 & 44.4  \\
N50 & & 00:53:26 & $-$72:42:56 & 4.3 & 76.3  \\
N51 & & 00:52:40 & $-$73:26:29 & 1.9 & 33.7 \\
N63 & & 00:58:17 & $-$72:38:57 & 1.3 & 23.1 \\
N66 & & 00:59:06 & $-$72:10:44 & 3.6 & 63.9 \\
N71 & & 01:00:59 & $-$71:35:30 & 0.2 & 3.55 \\
N76 & & 01:03:32 & $-$72:03:16 & 3.1 & 55.0 \\
N78 & & 01:05:18 & $-$71:59:53 & 2.6 & 46.1 \\
N80 & & 01:08:13 & $-$72:00:06 & 2.2 & 39.0 \\
N84 & & 01:14:56 & $-$73:17:51 & 5.7 & 101 \\
N90 & & 01:29:27 & $-$73:33:10 & 1.7 & 30.2 \\
\enddata
\tablenotetext{a}{Radii were selected such that they enclose 90\% of the H$\alpha$ emission of the sources. Radius in pc is calculated assuming distances of $D = 50$ kpc to the LMC and $D = 61$ kpc to the SMC.}  
\label{tab:sample}
\end{deluxetable*}

\begin{figure*}
\begin{center}
\includegraphics[width=\textwidth]{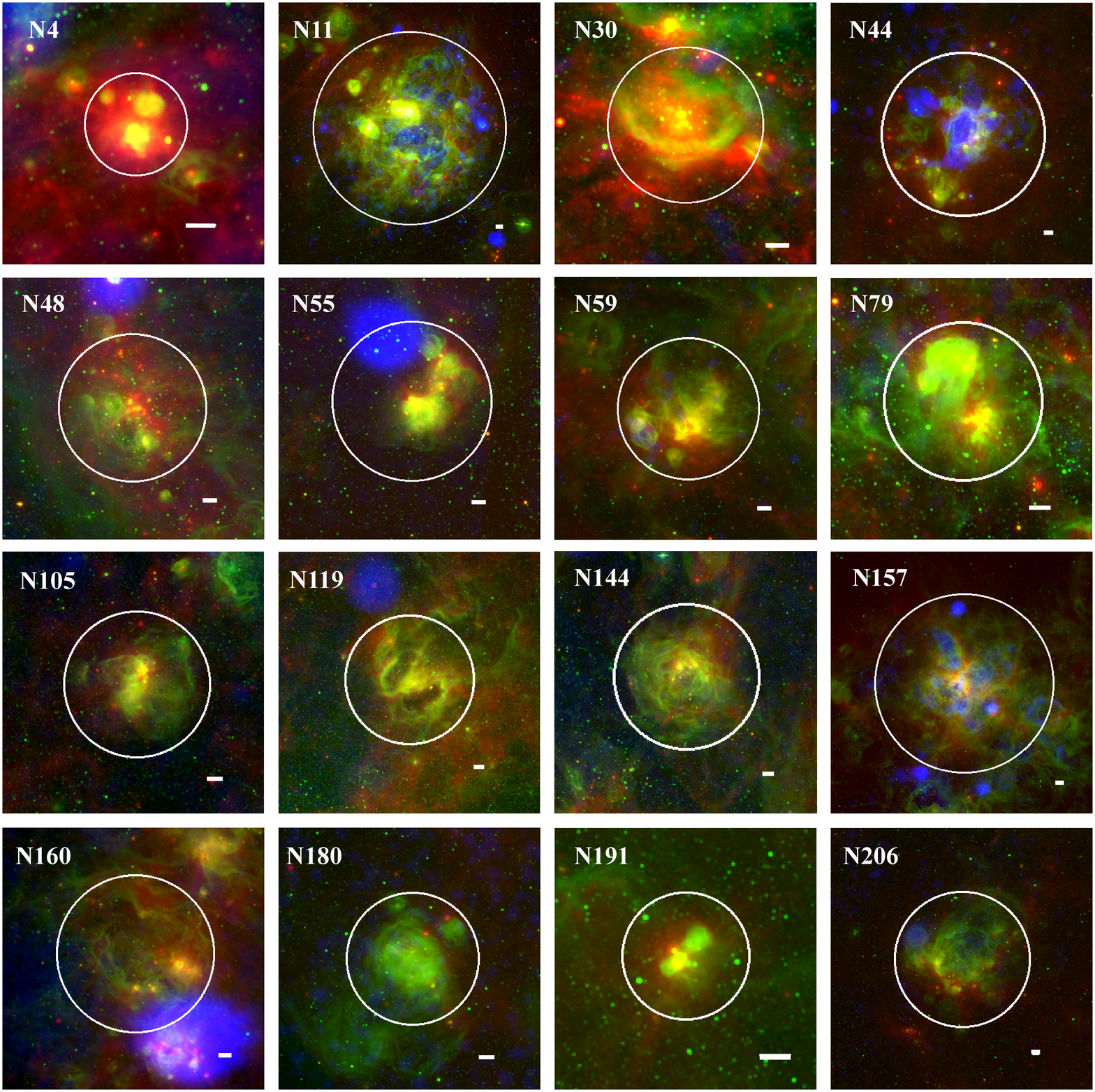}
\end{center}
\caption{Three-color images of the sixteen LMC \hii\ regions analyzed: MIPS 24$\mu$m (red), H$\alpha$ (green), and 0.2--2.1 keV X-rays (blue). White circles denote apertures used when calculating integrated pressures of the regions. The radius of each region was defined as the aperture which contained 90\% of the total H-$\alpha$ flux. We opted for this phenomenological definition of the radii to reduce the systematic uncertainties between sources. White bars in the bottom right of images have lengths of 1\arcmin\ $\approx$14.5 pc (assuming a distance of 50 kpc to the LMC). North is up, East is left.}
\label{fig:LMCthreecolor}
\end{figure*}

\begin{figure*}
\begin{center}
\includegraphics[width=\textwidth]{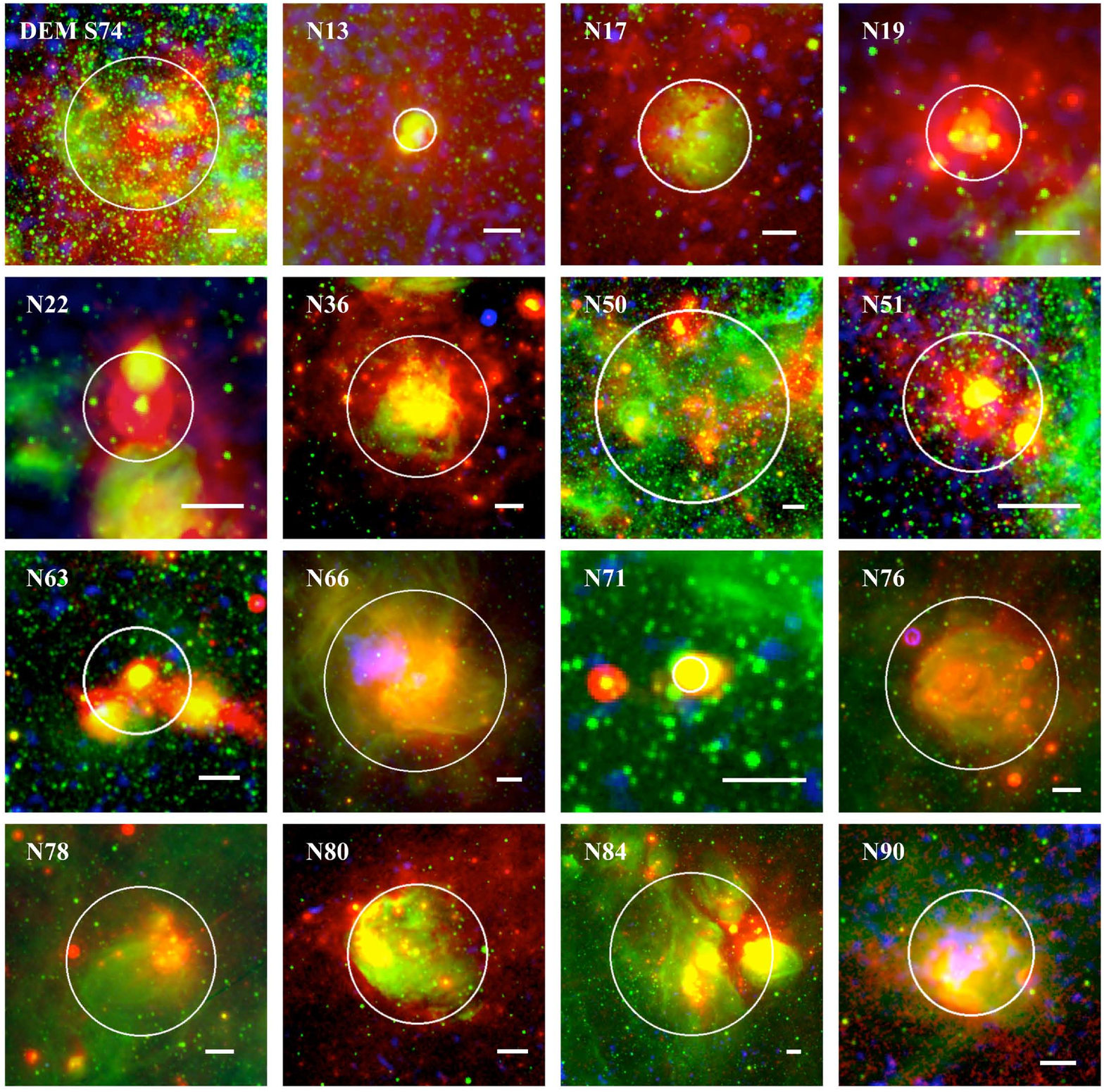}
\end{center}
\caption{Three-color images of the sixteen SMC \hii\ regions analyzed: MIPS 24 $\mu$m (red), H$\alpha$ (green) and 0.5--2.1 keV X-rays (blue). White circles denote apertures used when calculating integrated pressures of the regions. The radius of each region was defined as the aperture which contained 90\% of the total H-$\alpha$ flux. We opted for this phenomenological definition of the radii to reduce the systematic uncertainties between sources. White scale bars have lengths of 1\arcmin\  $\approx$ 17.7 pc (assuming a distance of 61 kpc to the SMC). North is up, East is left.}
\label{fig:SMCthreecolor}
\end{figure*}

Our final sample of \hii\ regions are listed in Table~\ref{tab:sample}, and Figures~\ref{fig:LMCthreecolor} and~\ref{fig:SMCthreecolor} shows the three-color images of the LMC and SMC \hii\ regions, respectively. We note that although our sample spans a range of parameter space (e.g., two orders of magnitude in radius and in ionizing photon fluxes $S$), the \hii\ regions we have selected represent the brightest in the Magellanic Clouds in H$\alpha$ and at 24 $\mu$m. 

We utilize published UBV photometry of 624 LMC star clusters \cite{bica96} to assess upper limits on the cluster ages and lower limits on star cluster masses powering our sample. Within the radii of the LMC \hii\ regions, we found 1--8 star clusters from the Bica sample. To estimate the cluster ages, we compare the extinction-corrected UBV colors of the enclosed star clusters to the colors output from Starburst99 simulations \citep{starburst99} of a star cluster of $10^{6} M_{\sun}$ which underwent an instantaneous burst of star formation. For this analysis, we adopt a color excess $E(B-V)=0.06$, the foreground reddening in the direction of the LMC \citep{oest95}. This value is almost certainly an underestimate and represents the minimum reddening toward our clusters (for example, the reddening in R136 is $E(B-V)=0.3-0.6$) and neglects local extinction. Based on the clusters' UBV colors, we find upper limit ages of $\sim$3--15 Myr; greater extinction toward the clusters would yield younger ages. Additionally, we estimate the lower limit of the star cluster masses by normalizing $10^{6} M_{\sun}$ by the ratio of the V-band luminosities of our clusters with those of the simulated clusters at their respective ages. We find cluster masses of $\sim$300--$3 \times 10^{4} M_{\sun}$.
 
As relatively bright and evolved sources, the dynamical properties of our sample may differ from more dim \hii\ regions (those powered by smaller star clusters) and \hii\ regions which are much younger or older. For our analyses, we employed data at several wavelengths; a brief description of these data is given below. Throughout this paper, we assume a distance $D$ of 50 kpc to the LMC \citep{pie13} and of 61 kpc to the SMC \citep{hilditch05}.

\subsection{Optical}

To illustrate the \hii\ regions' structure, we show the H$\alpha$ emission of the 32 sources in Figures~\ref{fig:LMCthreecolor} and~\ref{fig:SMCthreecolor}. We used the narrow-band image (at 6563$\mbox{\AA}$, with 30$\mbox{\AA}$ full-width half max) that was taken with the University of Michigan/CTIO 61-cm Curtis Schmidt Telescope at CTIO as part of the Magellanic Cloud Emission Line Survey (MCELS: \citealt{smith}). The total integration time was 600 s, and the reduced image has a resolution of 2\arcsec pixel$^{-1}$. 

To estimate the H$\alpha$ luminosity of our SMC sources within the radii given in Table~\ref{tab:sample}, we used the flux-calibrated, continuum-subtracted MCELS data. As the flux calibrated MCELS data of the LMC is not yet available, we employed the Southern H$\alpha$ Sky Survey Atlas (SHASSA), a robotic wide-angle survey of declinations $\delta = +15^{\circ}$ to $-90^{\circ}$ \citep{gaustad}, to measure H$\alpha$ luminosities of our LMC \hii\ regions. SHASSA data were taken using a CCD with a 52-mm focal length Canon lens at f/1.6. This setup enabled a large field of view ($13^{\circ} \times 13^{\circ}$) and a spatial resolution of 47.64\arcsec\ pixel$^{-1}$. The total integration time for the LMC exposure was $\approx$21 minutes.

\subsection{Infrared}

Infrared images of the LMC were obtained through the {\it Spitzer} Space Telescope Legacy program Surveying the Agents of Galaxy Evolution (SAGE: \citealt{meixner}). The survey covered an area of $\sim$7 $\times$ 7 degrees of the LMC with the Infrared Array Camera (IRAC; \citealt{fazio}) and the Multiband Imaging Photometer (MIPS; \citealt{rieke}). Images were taken in all bands of IRAC (3.6, 4.5, 5.8, and 7.9 $\mu$m) and of MIPS (24, 70, and 160 $\mu$m) at two epochs in 2005. For our analyses, we used the combined mosaics of both epochs with 1.2\arcsec pixel$^{-1}$ in the 3.6 and 7.9 $\mu$m IRAC images and 2.49\arcsec pixel$^{-1}$ and 4.8\arcsec pixel$^{-1}$ in the MIPS 24 $\mu$m and 70 $\mu$m, respectively. 

The SMC was also surveyed by {\it Spitzer} with the Legacy program Surveying the Agents of Galaxy Evolution in the Tidally Stripped, Low Metallicity Small Magellanic Cloud (SAGE-SMC: \citealt{gordon11}). This project mapped the full SMC ($\sim$30 deg$^2$) with IRAC and MIPS and built on the pathfinder program, the Spitzer Survey of the Small Magellanic Cloud (S$^{3}$MC: \citealt{bolatto07}), which surveyed the inner $\sim$3 deg$^{2}$ of the SMC. SAGE-SMC observations were taken at two epochs in 2007--2008, and we employed the combined mosaics from both epochs (plus the S$^{3}$MC data). 

\subsection{Radio}

The LMC and SMC were observed with the Australian Telescope Compact Array (ATCA) as part of 4.8-GHz and 8.64-GHz surveys \citep{dickel,dickel10}. These programs had identical observational setups, using two array configurations that provided 19 antenna spacings, and the ATCA observations were combined with the Parkes 64-m telescope data of \cite{haynes} to account for extended structure missed by the interferometric observations. For our analyses, we utilized the resulting ATCA$+$Parkes 8.64 GHz (3.5-cm) images of the LMC and SMC, which had Gaussian beams of FWHM 22\arcsec\ and an average rms noise level of 0.5 mJy beam$^{-1}$. 

\subsection{X-ray} 

Given the large extent of the LMC, {\it Chandra} and {\it XMM-Newton} have not observed the majority of that galaxy. Thus, for our X-ray analyses of the 16 LMC \hii\ regions, we use archival data from {\it ROSAT}, the R\"{o}ntgen Satellite. The LMC was observed via pointed observations and the all-sky survey of the ROSAT Position Sensitive Proportional Counter (PSPC) over its lifetime (e.g., \citealt{snowden}). The ROSAT PSPC had modest spectral resolution (with $\Delta E/E \sim 0.5$) and spatial resolution ($\sim$25\arcsec) over the energy range of 0.1--2.4 keV, with $\sim2^{\circ}$ field of view. Table~\ref{tab:xrayobslog} lists the archival PSPC observations we utilized in our analyses of our sample. All the LMC \hii\ regions except for N191 were observed in pointed observations from 1991--1993 with exposures ranging from $\sim$4000--45000 s. Some of these observations were presented and discussed originally in \cite{dunne01}.

\begin{deluxetable*}{lccc} 
\tablecolumns{4}
\tablewidth{0pt} \tablecaption{X-ray Observation Log} 
\tablehead{\colhead{Source} & \colhead{Obs Date} & \colhead{Archive Number} & \colhead{Integration (ks)}}
\startdata
\cutinhead{LMC Sources: ROSAT PSPC Observations} 
N4 & July 1993 & rp500263n00 & 12.7 \\
N11 & November 1992 & rp900320n00 & 17.6 \\
N30 & February 1992 & rp500052a01 & 8.0 \\
N44 & March 1992 & rp500093n00 & 8.7 \\
N48 & October 1991 & rp200692n00 & 44.7 \\
N55 & October 1991 & rp200692n00 & 44.7 \\
N59 & December 1993 & rp900533n00 & 1.6 \\
N79 & October 1993 & rp500258n00 & 12.7 \\
N105 & April 1992 & rp500037n00 & 6.8 \\
N119 & June 1993 & rp500138a02 & 14.6 \\
N144 & June 1993 & rp500138a02 & 14.6 \\
30 Dor & April 1992 & rp500131n00 & 16.0 \\
N160 & April 1992 & rp500131n00 & 16.0 \\
N180 & October 1993 & rp500259n00 & 4.0 \\
N191\tablenotemark{a} & -- & -- & -- \\
N206 & December 1993 & rp300335n00 & 11.3 \\
\cutinhead{SMC Sources: XMM-Newton or Chandra Observations} 
DEM~S74 & November 2009 & 0601211401 & 46.8 \\
N13 & October 2009 & 0601211301 & 32.7 \\
N17 & October 2009 & 0601211301 & 32.7 \\
N19 & March 2007 & 0403970301 & 39.1 \\
N22 & October 2000 & 0110000101 & 28.0 \\
N36 & March 2010 & 0656780201 & 12.8 \\
N50 & December 2003 & 0157960201 & 14.8 \\
N51 & April 2007 & 0404680301 & 20.4 \\
N63 & October 2009 & 0601211601 & 32.3 \\
N66\tablenotemark{b} & May 2001 & 1881 & 99.9 \\
N71 & June 2007 & 0501470101 & 16.1 \\
N76\tablenotemark{b} & March 2000--Jan 2009\tablenotemark{c} & 52 Observations\tablenotemark{c} & 471.0 \\
N78\tablenotemark{b} & Dec 2000--Feb 2009\tablenotemark{d} & 36 Observations\tablenotemark{d} & 297.6 \\
N80 & November 2009 & 0601211901 & 31.6 \\
N84 & March 2006 & 0311590601 & 11.3 \\
N90 & April 2010 & 0602520301 & 96.3 \\
\enddata
\tablenotetext{a}{N191 is not in any pointed PSPC observations, so we exclude it from our hot gas pressure analyses.} 
\tablenotetext{b}{For these sources, we analyze the {\it Chandra} ACIS observations instead of the {\it XMM-Newton} data because the {\it Chandra} observations have longer integrations.}
\tablenotetext{c}{N76 is in the field of the {\it Chandra} calibration source, SNR 1E 0102$-$7219, and has been observed repeatedly over {\it Chandra}'s lifetime. For our analysis of N76, we have merged 52 ACIS-I observations together with the following ObsIDs: 136, 140, 420, 439, 440, 444, 445, 1313, 1314, 1315, 1316, 1317, 1529, 1542, 1543, 2837, 2839, 2842, 2863, 3532, 3537, 3538, 3539, 3540, 5137, 5138, 5139, 5140, 5144, 5147, 5148, 5149, 5150, 5151, 5154, 6050, 6051, 6052, 6053, 6054, 6057, 6060, 6747, 6748, 6749, 6750, 6751, 6754, 6757, 8361, 8363, 10652.}
\tablenotetext{d}{N78 is in the field of the {\it Chandra} calibration source, SNR 1E 0102$-$7219, and has been observed repeatedly over {\it Chandra}'s lifetime. For our analysis of N78, we have merged 36 ACIS-I observations together with the following ObsIDs: 1527, 1528, 1533, 1534, 1535, 1536, 1537, 1544, 1783, 1784, 1785, 2840, 2841, 2858, 2859, 2860, 2861, 2864, 3527, 3528, 3529, 3530, 3531, 3541, 5145, 5152, 5153, 6055, 6056, 6060, 6753, 6755, 6757, 8362, 9691, 10650.}
\label{tab:xrayobslog}
\end{deluxetable*}

The SMC was surveyed by {\it XMM-Newton} between May 2009 and March 2010 \citep{haberl12}. We exploit data from this campaign as well as from pointed {\it XMM-Newton} observations for 13 of the 16 SMC \hii\ regions. For the other three SMC sources (N66, N76, and N78), we use deep {\it Chandra} ACIS-I observations. N66 was targeted in a 99.9 ks ACIS-I observation \citep{naze02,naze03}. N76 and N78 are in the field of a {\it Chandra} calibration source, the supernova remnant 1E 0102$-$7219, so they have been observed repeatedly since the launch of {\it Chandra} in 1999. We searched these calibration data and merged all the observations where the {\it Chandra} chip array imaged the full diameter of the sources: 52 observations for N76, and 36 observations for N78. 
 
\clearpage
 
\section{Methodology} \label{sec:method}

We follow the same methodology as in our 30 Dor pressure analysis \citep{lopez11} with only a few exceptions, described below. Instead of calculating spatially-resolved pressure components for the sources, we determine the average pressures integrated over the radii listed in Table~\ref{tab:sample}. Thus, these pressure components are those ``felt'' within the \hii\ shells. We describe the uncertainties associated with the calculation of each term in Section~\ref{sec:uncertainty}.

To select the radius of each region, we produced surface brightness profiles of their H-$\alpha$ emission, and we determined the apertures which contained 90\% of the total H-$\alpha$ fluxes. We opted for this phenomenological definition of the radii to reduce the systematic uncertainties between sources. As seen in Figures~\ref{fig:LMCthreecolor} and~\ref{fig:SMCthreecolor}, the \hii\ regions are quite complex, and the calculations below are simple and aimed to describe the general properties of these sources. 

\subsection{Direct Radiation Pressure} \label{sec:Pdir} 

The light output by stars produces a direct radiation pressure that is associated with the photons' energy and momentum. The resulting radiation pressure $P_{{\rm rad}}$ at some position within the \hii\ region is related to the bolometric luminosity of each star $L_{\rm bol}$ and the distance $r$ the light traveled to reach that point:

\begin{equation}
P_{{\rm rad}}= \sum \frac{L_{\rm bol}}{4 \pi r^{2} c},
\label{eq:Pdir}
\end{equation}

\noindent
where the summation is over all the stars in the region. The volume-averaged direct radiation pressure $P_{\rm dir}$ is then

\begin{equation}
P_{\rm dir} = \frac{1}{V} \int P_{\rm rad} dV = \frac{3}{4 \pi R^{3}} \int_{0}^{R} \frac{L_{\rm bol}}{c} dr = \frac{3 L_{\rm bol}}{4 \pi R^{2} c},
\label{eq:PdirV}
\end{equation}

\noindent
where $V$ is the total volume within the \hii\ region shell and $R$ is the \hii\ region radius. 

The above equation is the formal definition of radiation pressure (i.e., it is the trace of the radiation pressure tensor). We note that radiation pressure and radiation force do not always follow the same simple relationship as e.g., gas pressure and force, where the force is the negative gradient of pressure. In particular, \cite{pellegrini11} point out that in a relatively transparent medium (such as the interior of an \hii\ region), it is possible for the radiation pressure to exceed the gas pressure while the local force exerted on matter by the radiation is smaller than the force exerted by gas pressure. However, at the \hii\ shells where the gas is optically thick to stellar radiation, radiation force and pressure follow the same relationship as gas force and pressure, and the radiation pressure defined by Equation~\ref{eq:Pdir} is the relevant quantity to consider. 

To obtain $L_{\rm bol}$ of the stars in our 30 Dor analyses, we employed UBV photometry of R136 from \cite{mal94} using HST Planetary Camera observations, and the ground-based data of \cite{parker1} and \cite{selman05} to account for stars outside R136. While several large-scale optical surveys of the LMC have now been done and include UBV photometry (e.g., \citealt{massey02,zaritsky04}), these data do not resolve the crowded regions of young star clusters, and they focus principally on the (uncrowded) field population. 

An alternative means to estimate the bolometric luminosities of the star clusters is using the extinction-corrected H$\alpha$ luminosities of the \hii\ regions. From \cite{kennicutt12}, for a stellar population that fully samples the initial mass function (IMF) and the stellar age distribution, the bolometric luminosity $L_{\rm bol,IMF}$ is related to the extinction-corrected H$\alpha$ luminosity $L_{\rm H\alpha}$ by the expression $L_{\rm bol,IMF} = 138 L_{\rm H\alpha}$. We use the SHASSA and MCELS data to estimate the observed H$\alpha$ luminosities $L_{\rm H\alpha,obs}$ within the radii listed in Table~\ref{tab:sample}. 

To correct for extinction, we employ the reddening maps of the LMC and SMC presented in \cite{has11}, from the third phase of the Optical Gravitational Lensing Experiment (OGLE III). These authors used observations of red clump and RR Lyrae stars to derive spatially-resolved extinction estimates (with typical subfield sizes of 4.5\arcmin $\times$4.5\arcmin) across the LMC and SMC, and these data are publicly available through the German Astrophysical Virtual Observatory (GAVO) interface\footnote{http://dc.zah.uni-heidelberg.de/mcx}. Using GAVO, we obtained the mean extinction in the B- and V-bands, $A_{\rm B}$ and $A_{\rm V}$, respectively. In the cases when the \hii\ region radii included multiple subfields of the OGLE extinction measurements, we calculated the average $A_{\rm B}$ and $A_{\rm V}$ over that aperture. Then, we use the color excess $E(B-V) \equiv A_{\rm B} - A_{\rm V}$ to compute $A_{\rm H\alpha}$, the extinction at the wavelength $\lambda$ of the H$\alpha$ line, given

\begin{equation}
A_{\rm H\alpha} = k(H_{\rm H\alpha}) E(B-V), 
\end{equation}

\noindent
where $k(H_{\rm H\alpha}$) is the value of the extinction curve at the wavelength of the H$\alpha$ line. \cite{calzetti00} derive the best-fit expression for $k(\lambda$) at optical wavelengths as

\begin{equation}
k(\lambda) = 2.659 (-2.156 + 1.509/ \lambda - 0.198/  \lambda^{2} + 0.011/ \lambda^{3}) + R_{\rm V}
\end{equation}

\noindent
where $R_{\rm V} = A_{\rm V} / E(B-V)$. We adopt the standard $R_{\rm V} = 3.1$, which \cite{gordon03} demonstrate to be valid in the optical for the LMC and SMC, and we find $k(H\alpha) =$ 2.362. Finally, the extinction-corrected H$\alpha$ luminosity $L_{\rm H\alpha}$ is

\begin{equation}
L_{\rm H\alpha} = L_{\rm H\alpha,obs} 10^{0.4 A_{\rm H\alpha}}
\end{equation} 

\noindent
The parameters associated with these calculations, including the intrinsic H$\alpha$ luminosities and corresponding $L_{\rm bol,IMF}$ of the 32 \hii\ regions, are listed in Table~\ref{tab:extinction}. The extinction-corrected H$\alpha$ luminosities are typically 10--20\% greater than the observed H$\alpha$ luminosities. We note that local reddening and extinction may be greater than the average values obtained in the OGLE III maps, and thus the bolometric luminosities of the star clusters may be greater. However, even if the local extinction is a factor of ten larger, the direct radiation pressure will still be dynamically insignificant, as seen in the results given in Section~\ref{sec:results}.

\begin{deluxetable*}{lccccccc}
\tablecolumns{8}
\tablewidth{0pt} \tablecaption{Parameters to Estimate Extinction Correction} 
\tablehead{\colhead{Source} & \colhead{$A_{\rm B}$} & \colhead{$A_{\rm V}$} & \colhead{$A_{\rm H\alpha}$} & \colhead{log $L_{\rm H\alpha,obs}$\tablenotemark{a}} & \colhead{log $L_{\rm H\alpha}$\tablenotemark{b}} & \colhead{log $L_{\rm bol,IMF}$\tablenotemark{c}} & \colhead{log $S$} \\
\colhead{} & \colhead{} & \colhead{} & \colhead{} & \colhead{(erg s$^{-1}$)} &  \colhead{(erg s$^{-1}$)} & \colhead{(erg s$^{-1}$)} & \colhead{(ph s$^{-1}$)}}
\startdata
\cutinhead{LMC Sources}
N4 & 0.31 & 0.24 & 0.17 & 37.1 & 37.2 & 39.4 & 49.2 \\
N11 & 0.08 & 0.06 & 0.05 & 38.9 & 39.0 & 41.1 & 51.0 \\ 
N30 & 0.26 & 0.20 & 0.14 & 37.7 & 37.8 & 39.9 & 49.7 \\
N44 & 0.28 & 0.22 & 0.14 & 38.5 & 38.6 & 40.7 & 50.6 \\
N48 & 0.19 & 0.14 & 0.12 & 37.8 & 37.9 & 40.0 & 49.9 \\ 
N55 & 0.30 & 0.23 & 0.17 & 38.0 & 38.0 & 40.2 & 50.0 \\
N59 & 0.36 & 0.28 & 0.19 & 38.4 & 38.5 & 40.6 & 50.5 \\
N79 & 0.40 & 0.30 & 0.24 & 38.1 & 38.2 & 40.4 & 50.2 \\
N105 & 0.20 & 0.15 & 0.12 & 38.1 & 38.2 & 40.3 & 50.1 \\
N119 & 0.20 & 0.15 & 0.12 & 38.5 & 38.5 & 40.7 & 50.5 \\
N144 & 0.35 & 0.27 & 0.19 & 38.4 & 38.4 & 40.6 & 50.4 \\
N157 & 0.76 & 0.59 & 0.40 & 39.5 & 39.7 & 41.8 & 51.7 \\
N160 & 0.57 & 0.44 & 0.31 & 38.9 & 39.0 & 41.1 & 51.0 \\
N180 & 0.36 & 0.28 & 0.19 & 38.0 & 38.1 & 40.2 & 50.1 \\
N191 & 0.18 & 0.13 & 0.12 & 37.0 & 37.0 & 39.2 & 49.0 \\
N206 & 0.30 & 0.23 & 0.17 & 38.5 & 38.5 & 40.7 & 50.5 \\
\cutinhead{SMC Sources}
DEM~S74 & 0.16 & 0.12 & 0.09 & 37.1 & 37.1 & 39.3 & 49.1 \\
N13 & 0.25 & 0.19 & 0.14 & 37.0 & 37.1 & 39.2 & 49.0 \\
N17 & 0.21 & 0.16 & 0.12 & 37.1 & 37.2 & 39.3 & 49.1 \\
N19 & 0.25 & 0.19 & 0.14 & 36.7 & 36.8 & 38.9 & 48.8 \\
N22 & 0.27 & 0.21 & 0.14 & 37.0 & 37.1 & 39.2 & 49.1 \\
N36 & 0.24 & 0.18 & 0.14 & 37.8 & 37.9 & 40.0 & 49.9 \\
N50 & 0.19 & 0.14 & 0.12 & 37.8 & 37.8 & 39.9 & 49.8 \\
N51 & 0.15 & 0.12 & 0.08 & 36.8 & 36.8 & 39.0 & 48.8 \\
N63 & 0.22 & 0.17 & 0.12 & 37.0 & 37.0 & 39.1 & 49.0 \\
N66 & 0.08 & 0.06 & 0.05 & 38.6 & 38.6 & 40.8 & 50.6 \\
N71 & 0.11 & 0.09 & 0.05 & 36.2 & 36.3 & 38.4 & 48.2 \\
N76 & 0.09 & 0.07 & 0.05 & 38.0 & 38.0 & 40.2 & 50.0 \\
N78 & 0.13 & 0.10 & 0.07 & 37.7 & 37.7 & 39.9 & 49.7 \\
N80 & 0.16 & 0.12 & 0.09 & 37.4 & 37.5 & 39.6 & 49.4 \\
N84 & 0.32 & 0.24 & 0.19 & 38.2 & 38.2 & 40.4 & 50.2 \\
N90 & 0.19 & 0.14 & 0.12 & 37.5 & 37.5 & 39.7 & 49.5 \\
\enddata
\tablenotetext{a}{Observed H$\alpha$ luminosity (i.e., without extinction correction).} 
\tablenotetext{b}{Intrinsic H$\alpha$ luminosity (i.e., with extinction correction).} 
\tablenotetext{c}{$L_{\rm bol,IMF}$ is the bolometric luminosity estimated based on the intrinsic H$\alpha$ luminosity assuming a fully-sampled IMF.}
\tablenotetext{d}{$S$ is the ionizing photon rate, as calculated using $L_{\rm H\alpha}$ and Equation~\ref{eq:S}.}
\label{tab:extinction}
\end{deluxetable*}

One issue related to the above estimates of $L_{\rm bol,IMF}$ is the star formation history. While both the H$\alpha$ and bolometric luminosity of an actively star-forming region are dominated by massive stars with lifetimes $\ls 5$ Myr, the bolometric luminosity also contains a non-negligible contribution from longer-lived stars. The implication is that the ratio of H$\alpha$ to bolometric luminosity of a stellar population evolves with time. The relation $L_{\rm bol,IMF} = 138 L_{\rm H\alpha}$ is appropriate for a population with a continuous star formation history over 100 Myr, but for a nearly coeval stellar population as in our star clusters, the H$\alpha$ to bolometric ratio will start out somewhat larger than Kennicutt \& Evans value, then decline below it over a timescale of $\sim 5$ Myr. Thus, depending on the age of the stellar population, $L_{\rm bol,IMF}$ can be either an underestimate or an overestimate. Given that our stellar sources are bright H~\textsc{ii} regions and thus the stars are likely to be young, the latter seems more likely.

We also note uncertainty related to IMF sampling. Stellar populations with masses below $\sim 10^4$ $M_\odot$ do not fully sample the IMF, and this makes the H$\alpha$ to bolometric luminosity ratio vary stochastically \citep{cervino04,corbelli09,dasilva12}. Most of our clusters are near the edge of the stochastic regime. For a randomly selected cluster, the most common effect is to lower the H$\alpha$ luminosity relative to the bolometric luminosity; the expected magnitude of the effect is a factor of $\sim 3$ (e.g., Figure 7 of \citealt{corbelli09}). This will tend to make our $L_{\rm bol,IMF}$ an underestimate by this amount. However, the real effect is likely to be smaller, because our sample is not randomly selected. For a rare subset of clusters stochasticity has no effect or actually raises the H$\alpha$ to bolometric ratio compared to that of a fully-sampled IMF, and since our sample is partly selected based on H$\alpha$ luminosity, it is biased in favor of the inclusion of such clusters. It is not possible to model this effect quantitatively without knowing both the underlying distribution of cluster masses and the selection function used to construct the sample. Thus we restrict ourselves to noting that this effect probably introduces a factor of $\sim 2$ level uncertainty into $L_{\rm bol,IMF}$. In the remainder of this paper, we will use $L_{\rm bol,IMF} = L_{\rm bol}$ to calculate $P_{\rm dir}$.

\subsection{Dust-Processed Radiation Pressure} \label{sec:PIR}

The pressure of the dust-processed radiation field $P_{\rm IR}$ is related to the energy density of the radiation field absorbed by the dust, $u_{\nu}$ (i.e., assuming a steady state), 

\begin{equation}
P_{{\rm IR}} = \frac{1}{3} u_{\nu}.
\label{eq:PIR}
\end{equation}

\noindent
We follow the same procedure of \cite{lopez11} to estimate the energy density $u_{\nu}$ of the radiation absorbed by the dust in our sample. Specifically, we measure the flux densities $F_{\nu}$ in the IRAC and MIPS bands and compare them to the predictions of the dust models of \citealt{dl07} (hereafter DL07). The DL07 models determine the IR spectral energy distribution for a given dust content and radiation field intensity heating the dust. DL07 assume a mixture of carbonaceous grains and amorphous silicate grains that have a size distribution that reproduces the wavelength-dependent extinction in the local Milky Way (see \citealt{wein01}). In particular, polycyclic aromatic hydrocarbons (PAHs) contribute substantial flux at  $\sim$3--19 $\mu$m and are observed in normal and star-forming galaxies (e.g., \citealt{helou00,smith07}). 

To account for the different spatial resolutions of the IR images, we convolved the 3.6, 8, and 24 $\mu$m images with kernels to match the point-spread function of the 70 $\mu$m image using the convolution kernels of \cite{gordon}. Then, we measured the average flux densities $F_{\nu}$ at 8, 24, and 70 $\mu$m wavelengths in the apertures listed in Column 5 of Table~\ref{tab:sample}. We removed the contribution of starlight to the 8 and 24 $\mu$m fluxes using the 3.6 $\mu$m flux densities and the empirical relations

\begin{eqnarray}
F_{\nu}^{\rm{ns}}(8 \mu m) & = & F_{\nu}(8 \mu m) - 0.232F_{\nu}(3.6 \mu m) \\
F_{\nu}^{\rm{ns}}(24 \mu m) & = & F_{\nu}(24 \mu m) - 0.032F_{\nu}(3.6 \mu m)
\end{eqnarray}
 
\noindent
where $F_{\nu}^{\rm ns}$ is the non-stellar flux at the respective wavelengths. The coefficients 0.232 and 0.032 are given in \cite{helou}. 

\begin{figure}
\includegraphics[width=\columnwidth]{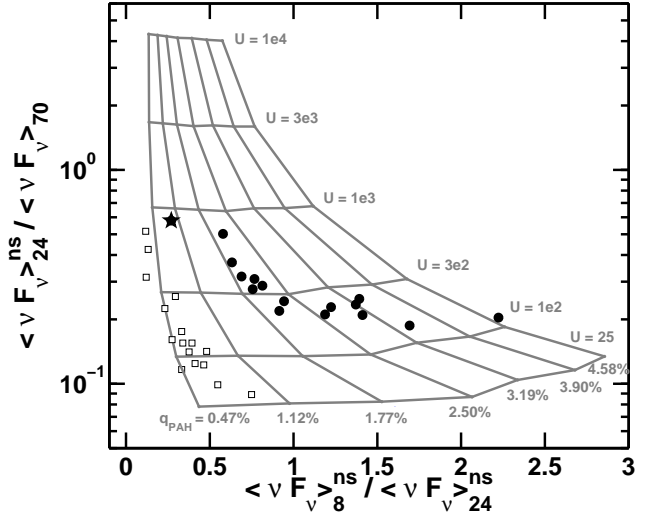}
\caption{Measured IR flux ratios for the 16 LMC \hii\ regions (filled black circles) and 16 SMC \hii\ regions (open squares) and the predicted flux ratios for different PAH mass fractions $q_{\rm PAH}$ and scaling $U$ of the energy density of the dust-processed radiation field (Equation~\ref{eq:u}) from \citealt{dl07}. The black star denotes the values for 30 Dor. We interpolate the grid of predicted flux ratios to obtain $q_{\rm PAH}$ and $U$ for each region listed in Table~\ref{table:PIRresults}.}
\label{fig:LMC_models}
\end{figure}

In Figure~\ref{fig:LMC_models}, we plot the resulting ratios $\langle \nu F_{\nu} \rangle_{{\rm 24}}^{{\rm ns}}/ \langle \nu F_{\nu} \rangle_{{\rm 70}}$ versus $\langle \nu F_{\nu} \rangle_{{\rm 8}}^{{\rm ns}}/ \langle \nu F_{\nu} \rangle_{{\rm 24}}^{{\rm ns}}$ measured for the 32 \hii\ regions. Additionally, we plot the \citealt{dl07} predictions for given values of $q_{{\rm PAH}}$, the fraction of dust mass in PAHs, and $U$, the dimensionless scale factor of energy density $u_{\nu}$ of radiation absorbed by the dust, where

\begin{equation}
u_{\nu} = U u_{\nu}^{{\rm IRSF}}.
\label{eq:u}
\end{equation}

\noindent
Here, $u_{\nu}^{{\rm IRSF}}$ is the energy density of the $h \nu < 13.6$ eV photons in the local ISM, 8.65 $\times$ 10$^{-13}$ erg cm$^{-3}$ \citep{mathis}.

The 32 \hii\ regions span a factor of $\sim$20 in $\langle \nu F_{\nu} \rangle_{{\rm 8}}^{{\rm ns}}/ \langle \nu F_{\nu} \rangle_{{\rm 24}}^{{\rm ns}}$, with the SMC \hii\ regions having systematically lower $\langle \nu F_{\nu} \rangle_{{\rm 8}}^{{\rm ns}}/ \langle \nu F_{\nu} \rangle_{{\rm 24}}^{{\rm ns}}$ than the LMC \hii\ regions. The LMC and SMC sources have a similar range of a factor of $\sim$6 in  $\langle \nu F_{\nu} \rangle_{{\rm 24}}^{{\rm ns}}/ \langle \nu F_{\nu} \rangle_{{\rm 70}}$. Broadly, the data follow a similar arc-like trend in the ratios as we found in the spatially-resolved regions of 30 Dor \citep{lopez11}. Errors in our flux ratios are $\approx$2.8\% from a $\approx$2\% uncertainty in the {\it Spitzer} photometry.

\begin{figure}
\includegraphics[width=\columnwidth]{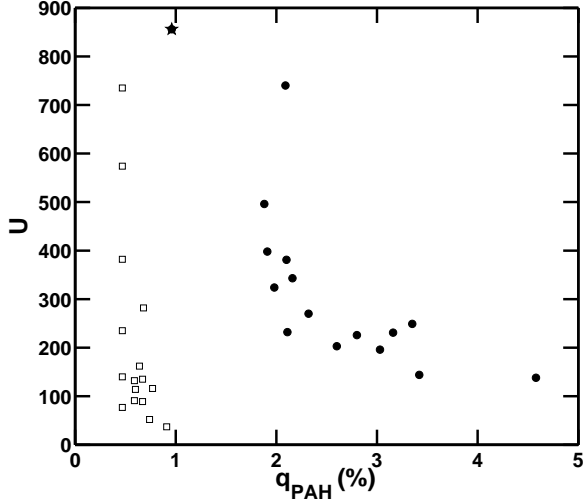}
\caption{Plot of $U$ versus PAH fraction $q_{\rm PAH}$ for the 16 LMC \hii\ regions (black circles) and 16 SMC \hii\ regions (open squares), as given by the interpolation of the grid in Figure~\ref{fig:LMC_models}. The numerical values for the two parameters are given in Table~\ref{table:PIRresults}, and the black star denotes the values for 30 Dor.}
\label{fig:LMC_UvsPAH}
\end{figure}

We interpolate the $U$-$q_{\rm PAH}$ grid using Delaunay triangulation, a technique appropriate for a non-uniform grid, to find the $U$ and $q_{{\rm PAH}}$ values for our regions. For the points that lay outside the grid, we translated them to $\langle \nu F_{\nu} \rangle_{{\rm 8}}^{{\rm ns}}/ \langle \nu F_{\nu} \rangle_{{\rm 24}}^{{\rm ns}}$ within the grid. Since the y-axis ratio largely determines $U$, this adjustment does not affect the pressure calculation for those sources. Figure~\ref{fig:LMC_UvsPAH} plots the interpolated values of $U$ versus $q_{{\rm PAH}}$; we also print the results in Table~\ref{table:PIRresults} so individual sources can be identified. We find that the $U$ values of the LMC and SMC \hii\ regions span a large range, with $U \approx 37$--856 (corresponding to $u_{\nu} \approx 3.2\times10^{-11}$--7.4$\times10^{-10}$ erg cm$^{-3}$), and several of the SMC sources have $U <100$. The PAH fractions of the SMC \hii\ regions (with $q_{\rm PAH} \ls$1\%) are suppressed relative to those of the LMC \hii\ regions (with $q_{\rm PAH} \gs$1\%). The smaller PAH fractions in the low metallicity SMC are consistent with the results of \cite{sandstrom12}, who find a deficiency of PAHs in the SMC based on observations with the {\it Spitzer} Infrared Spectrograph. Based on PAH band ratios in the IRS data, these authors suggest that this deficiency arises because SMC PAHs are smaller and more neutral than PAHs in higher metallicity galaxies. 

\begin{deluxetable}{lrrr}  
\tablecolumns{4}
\tablewidth{0pt} \tablecaption{Dust and Warm Gas Properties}
\tablehead{\colhead{Source} & \colhead{$U$} & \colhead{$q_{\rm PAH}$} & \colhead{$n_{\rm e}$ (cm$^{-3}$)}}
\startdata
\cutinhead{LMC Sources} 
N4 & 740 & 2.1 & 500 \\
N11 & 230 & 3.2 & 50 \\
N30 & 250 & 3.4 & 60 \\
N44 & 230 & 2.8 & 60 \\
N48 & 140 & $>$4.6 & 50 \\
N55 & 200 & 2.6 & 50 \\
N59 & 400 & 1.9 &  120 \\
N79 & 320 & 2.0 & 80 \\
N105 & 340 & 2.2 & 130 \\
N119 & 200 & 3.0 & 60 \\
N144 & 270 & 2.3 & 70 \\
30 Dor & 860 & 1.0 & 250  \\
N160 & 380 & 2.1 & 120 \\
N180 &  230 & 2.1 & 120 \\
N191 & 500 & 1.9 & 50 \\
N206 & 140 & 3.4 & 50 \\
\cutinhead{SMC Sources}
DEM~S74 & 40 & 0.9 & 30 \\
N13 & 280 & 0.7 & 260 \\
N17 & 120 & 0.8 & 70 \\
N19 & 140 & $<$0.5 & 160 \\
N22 & 740 & $<$0.5 & 160 \\
N36 & 80 & $<$0.5 & 60 \\
N50 & 50 & 0.7 & 20 \\
N51 & 140 & 0.7 & 30 \\
N63 & 90 & 0.7 & 60 \\
N66 & 380 & $<$0.5 & 100 \\
N71 & 240 & $<$0.5 & 330 \\
N76 & 130 & 0.6 & 70  \\
N78 & 570 & $<$0.5 & 70 \\
N80 & 90 & 0.6 & 50 \\
N84 & 160 & 0.6 & 30 \\
N90 & 110 & 0.6 & 50 \\
\enddata
\label{table:PIRresults}
\end{deluxetable}

Finally, we employ the interpolated $U$ values and Equations~\ref{eq:PIR} and~\ref{eq:u} to estimate the dust-processed radiation pressure $P_{\rm IR}$ in our 32 sources.

\subsection{Warm Ionized Gas Pressure} \label{sec:PHII}

The warm ionized gas pressure is given by the ideal gas law, $P_{\rm HII} = (n_{\rm e} + n_{\rm H} + n_{\rm He}) kT_{\rm HII}$, where $n_{{\rm e}}$, $n_{\rm H}$, and $n_{\rm He}$ are the electron, hydrogen, and helium number densities, respectively, and $T_{\rm HII}$ is temperature of the HII gas, which we assume to be the same for electrons and ions. If helium is singly ionized, then $n_{{\rm e}} + n_{\rm H} + n_{\rm He} \approx 2 n_{\rm e}$. If we adopt the temperature $T_{\rm HII}$ = 10$^{4}$ K, then the warm gas pressure is determined by the electron number density $n_{\rm e}$. One way to estimate $n_{\rm e}$ is via fine-structure line ratios in the IR (e.g., \citealt{indebetouw09}): since these lines have smaller excitation potentials than optical lines, they depend less on temperature and depend sensitively on the density \citep{osterbrock06}.

Here, we estimate $n_{\rm e}$ using an alternative means: by measuring the average flux density $F_{\nu}$ at 3.5~cm, where free-free emission dominates in \hii\ regions. For free-free emission, $n_{\rm e}$ is given by Eq.~5.14b of \cite{rybicki79}:

\begin{equation}
n_{\rm e} = \bigg( \frac{6.8 \times 10^{38} 4 \pi D^2 F_{\nu} T_{{\rm HII}}^{1/2}}{\bar{g}_{\rm ff} V} \bigg)^{1/2}, 
\label{eq:ne}
\end{equation}

\noindent
where $\bar{g}_{\rm ff}$ is the Gaunt factor and $D$ is the distance to the sources, and $V$ is the volume of the sources. If we set the Gaunt factor $\bar{g}_{\rm ff} = 1.2$, we derive the densities $n_{\rm e}$ listed in Table~\ref{table:PIRresults}. We find both the LMC and SMC \hii\ regions have moderate densities, with $n_{\rm e}\approx$ 22--500 cm$^{-3}$. 

\subsection{Hot Gas Pressure} \label{sec:Px}

The hot gas pressure is also given by an ideal gas law: $P_{\rm X} = 1.9 n_{\rm X} k T_{\rm X}$, where $n_{\rm X}$ is the electron density and $T_{\rm X}$ is the temperature of the X-ray emitting gas. The factor of 1.9 is derived assuming that He is doubly ionized and the He mass fraction is 0.3. Furthermore, we assume that the electrons and ions have reached equipartition, so that a single temperature describes both populations. To estimate $n_{\rm X}$ and $T_{\rm X}$, we model the bremsstrahlung emission at X-ray wavelengths of our sources using pointed {\it ROSAT} PSPC observations (for the LMC sources) and {\it Chandra} observations (for N66 in the SMC). The other \hii\ regions in the SMC are undetected by {\it XMM-Newton} and {\it Chandra}, and we use these data to set upper limits on hot gas pressure in those targets. In the analyses described below, we assume a filling factor $f_{\rm X}=1$ of the hot gas (i.e. that the hot gas occupies the full volume of our sources). For the purposes of measuring the large-scale dynamical role of the hot gas, the appropriate quantity is the volume-averaged pressure. We explain in detail why this approach is critical when assessing global dynamics in Appendix~\ref{app:hot gas}. 

\begin{figure*}
\includegraphics[width=\textwidth]{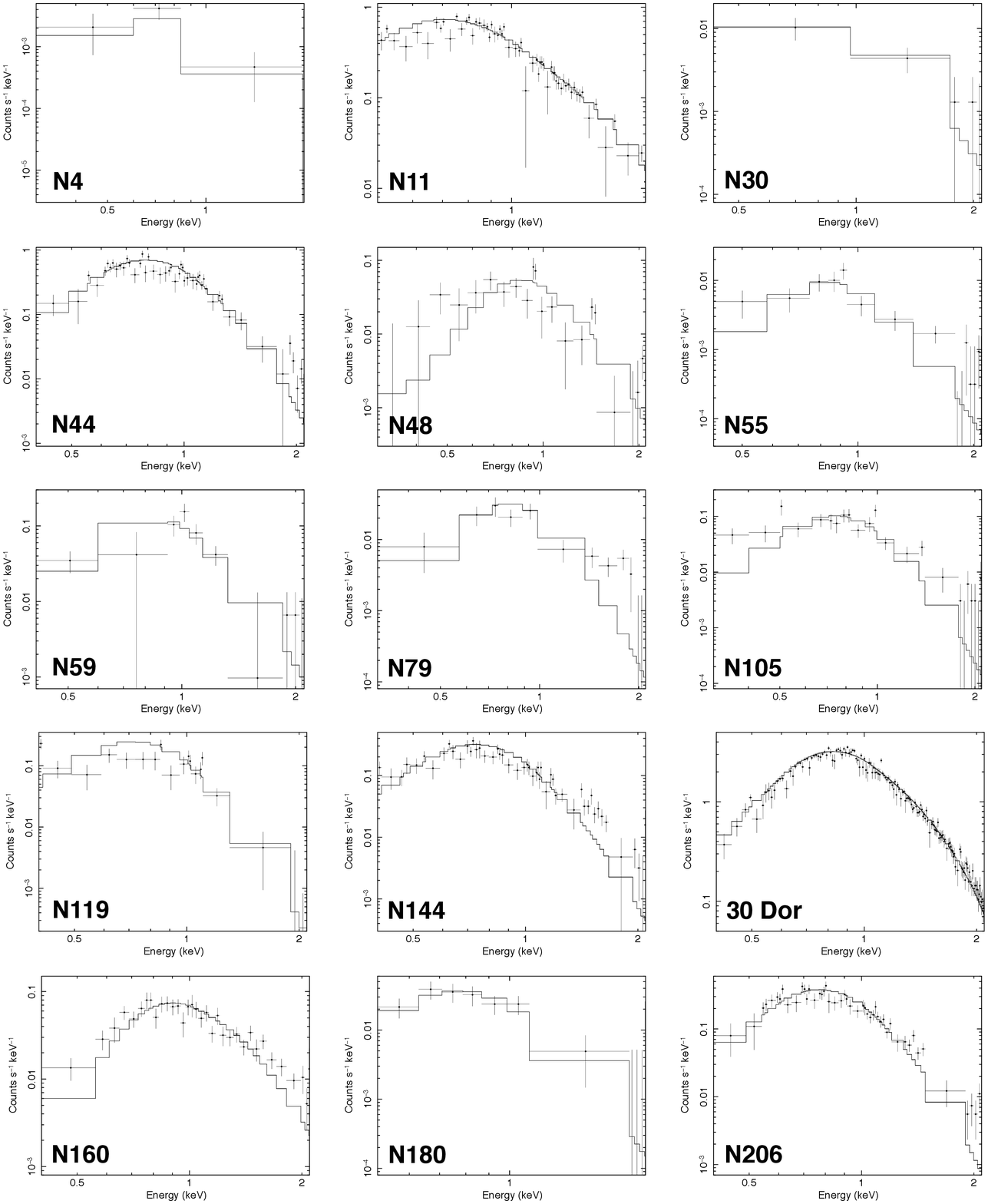}
\caption[Integrated ROSAT spectra and best-fit models for 15 LMC \hii\ regions]{Integrated background-subtracted ROSAT X-ray spectra for the 15 LMC \hii\ regions.}
\label{fig:ROSAT15}
\end{figure*} 

For the {\it ROSAT} analyses of the LMC \hii\ regions, we used {\sc ftools}, a software package for processing general and mission-specific FITS data \citep{blackburn95}, and {\sc xselect}, a command-line interface of {\sc ftools} for analysis of X-ray astrophysical data. We produced X-ray images of the sources (shown in blue in Figure~\ref{fig:LMCthreecolor}), and we extracted spectra from within the radii given in Table~\ref{tab:sample} as well as from background regions to subtract from the source spectra. Appropriate response matrices (files with probabilities that a photon of a given energy will produce an event in a given channel) and ancillary response files (which has information like effective area) were downloaded\footnote{Response matrices and ancillary response files are available via anonymous ftp at ftp://legacy.gsfc.nasa.gov/caldb/data/rosat/pspc/cpf/.} for each observation's date and detector. 

Resulting background-subtracted source spectra (shown in Figure~\ref{fig:ROSAT15}) were fit using XSPEC Version 12.4.0 \citep{arnaud96}. Spectra were modeled as an absorbed hot diffuse gas in collisional ionization equilibrium (CIE) using the XSPEC components {\it phabs} and {\it apec}. In these fits, we assumed a metallicity $Z \sim 0.5 Z_{\sun}$, the value measured in \hii\ regions in the LMC \citep{kd98}, and we adopted the solar abundances of \cite{asplund09}. In some sources (N11, 30 Dor, and N160), we found the addition of a power-law component was necessary in order to account for excess flux at energies $\gs$2 keV, a feature that is likely to be from non-thermal emission from supernova remnants or from point sources in the regions.

\begin{deluxetable*}{llccll} 
\tablecolumns{6}
\tablewidth{0pt} \tablecaption{X-ray Spectral Fit Results} 
\tablehead{\colhead{Source} & \colhead{$N_{\rm H}$} & \colhead{$kT_{\rm X}$} & \colhead{$n_{\rm X}$} & \colhead{log $L_{\rm X}$\tablenotemark{b}} & \colhead{$\chi^{2}$/d.o.f.} \\
\colhead{} & \colhead{($\times10^{21}$ cm$^{-2}$)} & \colhead{(keV)} & \colhead{(cm$^{-3}$)} & \colhead{(erg s$^{-1}$)} & \colhead{}}
\startdata
\cutinhead{LMC Sources}
N4 & 1.6\tablenotemark{a} & 0.15$\pm$0.04 & 0.28$\pm$0.27 & 34.1 & 13/9 \\ 
N11 & 1.9\tablenotemark{a} & 0.20$\pm$0.01 & 0.04$\pm$0.01 & 36.3 & 100/99 \\ 
N30 & 1.9\tablenotemark{a} & 0.67$\pm$0.30 & 0.27$\pm$0.09 & 34.6 & 20/52 \\ 
N44 & 6.0 & 0.22$\pm$0.07 & 0.12$\pm$0.07 & 37.0 & 156/107 \\ 
N48 & 4.7 & 0.54$\pm$0.41 & 0.03$\pm$0.02 & 35.6 & 135/123 \\ 
N55 & 1.2\tablenotemark{a} & 0.62$\pm$0.16 & 0.01$\pm$0.01 & 34.4 & 34/53 \\ 
N59 & 1.6\tablenotemark{a} & 0.63$\pm$0.13 & 0.04$\pm$0.02 & 35.6 & 19/54 \\
N79 & 1.6\tablenotemark{a} & 0.45$\pm$0.12 & 0.02$\pm$0.01 & 35.1 & 47/47 \\ 
N105 & 2.1\tablenotemark{a} & 0.25$\pm$0.03 & 0.09$\pm$0.04 & 35.6 & 68/74 \\ 
N119 & 2.1\tablenotemark{a} & 0.23$\pm$0.01 & 0.06$\pm$0.02 & 35.9 & 181/109 \\ 
N144 & 2.0\tablenotemark{a} & 0.25$\pm$0.01 & 0.07$\pm$0.02 & 36.0 & 166/115 \\ 
30 Dor & 3.0\tablenotemark{a} & 0.39$\pm$0.04 & 0.08$\pm$0.03 & 36.8 & 204/165 \\ 
N160 & 8.1 & 0.54$\pm$0.17 & 0.04$\pm$0.03 & 34.8 & 62/40 \\ 
N180 & 2.5\tablenotemark{a} & 0.30$\pm$0.06 & 0.06$\pm$0.03 & 35.2 & 11/31  \\ 
N191 & -- & -- & -- & -- & -- \\
N206 & 3.0 & 0.28$\pm$0.14 & 0.05$\pm$0.04 & 36.3 & 141/96 \\ 
\cutinhead{SMC Sources}
N66 & 3.3\tablenotemark{a} & 0.38$\pm$0.01 & 0.06$\pm$0.03 & 35.7 & 128/86 \\ 
\enddata
\label{table:PXresults}
\tablenotetext{a}{$N_{\rm H}$ was frozen to the weighted average value in the direction of the source, as obtained by the Leiden/Argentine/Bonn Survey of Galactic H{\sc i} from \cite{lab05}.} 
\tablenotetext{b}{X-ray luminosity of the thermal emission from the sources, corrected for absorption and in the 0.5--2.0 keV band.}
\end{deluxetable*}

\begin{figure}
\includegraphics[width=0.85\columnwidth,angle=-90]{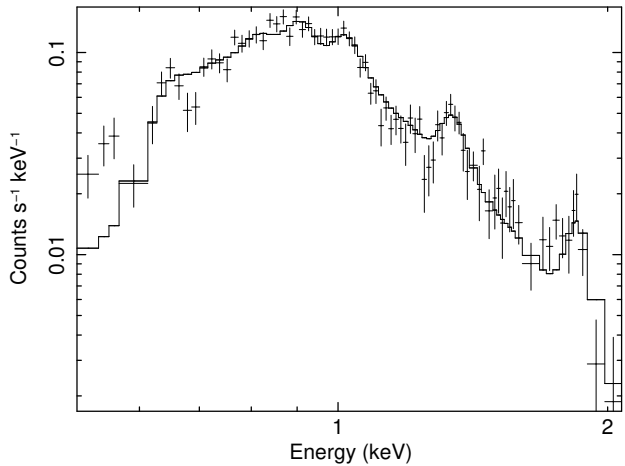}
\caption{Background-subtracted {\it Chandra} X-ray spectrum for the SMC \hii\ region N66. The best-fit model was an absorbed CIE plasma with enhanced abundances of O, Ne, and Si relative to the SMC metallicity of 0.2 $Z_{\sun}$. These enhanced abundances suggest the X-ray emission in N66 arises from a relatively young (a few thousand years old) supernova remnant.}
\label{fig:n66spectrum}
\end{figure} 

\begin{deluxetable*}{lcccclc}
\tablecolumns{7} \tablewidth{0pt}
\tablecaption{X-ray Upper Limits for SMC Sources} 
\tablehead{\colhead{Source} & \colhead{$N_{\rm H}$} & \colhead{Count Rate\tablenotemark{a}} & \colhead{$F_{\rm X,abs}$\tablenotemark{b}} & \colhead{$F_{\rm X, unabs}$\tablenotemark{c}} & \colhead{log $L_{\rm X}$\tablenotemark{d}} & \colhead{$n_{\rm X}$\tablenotemark{e}} \\
\colhead{} & \colhead{($\times10^{21}$ cm$^{-2}$)} & \colhead{(cts s$^{-1}$)} & \colhead{(erg cm$^{-2}$ s$^{-1}$)} & \colhead{(erg cm$^{-2}$ s$^{-1}$)} & \colhead{(erg s$^{-1}$)} & \colhead{(cm$^{-3}$)}} 
\startdata
DEM~S74 & 5.06 & 0.0293 & 1.8$\times10^{-13}$ & 4.6$\times10^{-12}$ & 36.3 & 0.37 \\
N13 & 3.58 & 0.0013 & 8.7$\times10^{-15}$ & 1.0$\times10^{-14}$ & 33.6 & 0.69 \\
N17 & 3.33 & 0.0078 & 5.3$\times10^{-14}$ & 5.2$\times10^{-13}$ & 35.4 & 0.31 \\
N19 & 4.76 & 0.0026 & 1.6$\times10^{-14}$ & 3.6$\times10^{-13}$ & 35.2 & 0.76 \\
N22 & 4.44 & 0.0025 & 1.6$\times10^{-14}$ & 3.0$\times10^{-13}$ & 35.1 & 0.52 \\
N36 & 5.02 & 0.0241 & 1.5$\times10^{-13}$ & 3.7$\times10^{-12}$ & 36.2 & 0.41 \\
N50 & 4.86 & 0.0532 & 3.3$\times10^{-13}$ & 7.7$\times10^{-12}$ & 36.5 & 0.24 \\
N51 & 4.41 & 0.0137 & 8.7$\times10^{-14}$ & 1.6$\times10^{-12}$ & 35.9 & 0.39 \\
N63 & 4.60 & 0.0065 & 4.1$\times10^{-14}$ & 8.3$\times10^{-13}$ & 35.6 & 0.55 \\
N71 & 2.49 & 0.0002 & 1.1$\times10^{-15}$ & 6.5$\times10^{-15}$ & 33.5 & 0.70 \\
N76 & 3.45 & 0.1821 & 2.9$\times10^{-12}$ & 3.1$\times10^{-11}$ & 37.1 & 0.46 \\
N78 & 3.49 & 0.0853 & 1.3$\times10^{-12}$ & 1.5$\times10^{-11}$ & 36.8 & 0.41 \\
N80 & 3.48 & 0.0173 & 1.2$\times10^{-13}$ & 1.3$\times10^{-12}$ & 35.8 & 0.25 \\
N84 & 3.52 & 0.2549 & 1.7$\times10^{-12}$ & 1.9$\times10^{-11}$ & 36.9 & 0.23 \\
N90 & 2.10 & 0.0194 & 1.4$\times10^{-13}$ & 6.4$\times10^{-13}$ & 35.5 & 0.26 \\
\enddata
\label{table:PXupperlimits}
\tablenotetext{a}{Count rate in the 0.5--8.0 keV band observed by {\it XMM-Newton} or {\it Chandra} within the radius of the \hii\ region.} 
\tablenotetext{b}{Upper limit on the absorbed flux from the source in the 0.5--10.0 keV band, as predicted by WebPIMMS based on the measured count rates.}
\tablenotetext{c}{Upper limit on the unabsorbed flux from the source in the 0.5--10.0 keV band, as predicted by WebPIMMS based on the measured count rates and $N_{\rm H}$.}
\tablenotetext{d}{Upper limit on the absorption-corrected X-ray luminosity in the 0.5--10.0 keV band.}
\tablenotetext{e}{Upper limit on $n_{\rm X}$, determined from the emission of a simulated $Z = 0.2 Z_{\sun}$, $kT_{\rm X} = 0.15$keV X-ray spectrum of a source with an X-ray flux equal to that listed in Column 5.}
\end{deluxetable*}

For the {\it Chandra} analysis of N66, we extracted a source spectrum using the CIAO command {\it specextract}; a background spectrum was obtained from a circular region of radius $\sim$50\arcsec\ offset $\sim$1\arcmin\ northeast of N66. The resulting background-subtracted spectrum (grouped to 25 counts per bin) is shown in Figure~\ref{fig:n66spectrum}. We first attempted to fit the spectrum with an absorbed hot diffuse gas in CIE as above (with XSPEC components {\it phabs} and {\it apec}) assuming a $Z = 0.2 Z_{\sun}$ metallicity plasma. The fit was statistically poor (with reduced chi-squared values of $\chi^{2}$/d.o.f. $=$ 317/90), with the greatest residuals around emission line features. Consequently, we considered an absorbed CIE plasma with varying abundances (with XSPEC components {\it phabs} and {\it vapec}). In this model, we let the abundances of elements in the spectrum (O, Ne, Mg, Si, and Fe) vary freely. The fit was dramatically improved (with $\chi^{2}$/d.o.f. $=$ 128/86) in this case. We found that the Mg and Fe abundances were consistent with those of the SMC, while O, Ne, and Si had enhanced abundances of $\sim$0.7 $Z_{\sun}$.  The elevated metallicity of the hot plasma is suggestive that the X-ray emission is from a relatively young (a few thousand years old) supernova remnant (SNR), and the enhanced abundances are signatures of reverse shock-heated ejecta. A young SNR in N66 has been identified previously as SNR B0057$-$724 based on its non-thermal radio emission  \citep{ye91}, its high-velocity H$\alpha$ emission \citep{chu88}, and its far-ultraviolet absorption lines \citep{danforth03}. 

The {\it ROSAT} and {\it Chandra} X-ray spectral fit results are given in Table~\ref{table:PXresults}, including the absorbing column density $N_{\rm H}$, the hot gas temperature $kT_{\rm X}$, the hot gas electron density $n_{\rm X}$, their associated 90\% confidence limits, and the reduced chi-squared for the fits, $\chi^{2}$/d.o.f. Hot gas temperatures were generally low, with $kT_{\rm X} \sim$ 0.15--0.6 keV. Comparing {\it ROSAT} results for 30 Dor to those from {\it Chandra} in \cite{lopez11}, we find that the integrated {\it Chandra} spectral fits gave temperatures a factor of $\sim$60\% above those given by {\it ROSAT}. This difference can be attributed to the fact that the {\it ROSAT} spectra were extracted from a much larger aperture than those from {\it Chandra}. Broadly, the X-ray luminosity $L_{\rm X}$ derived from our fits are consistent with previous X-ray studies of \hii\ regions in the LMC \citep{chu90,wang91,chu95}.

For the SMC \hii\ regions (except N66), we calculate upper limits on $P_{\rm X}$ based on the non-detections of these sources in {\it Chandra} (for N76 and N78) and {\it XMM-Newton} data. In particular, we measured the full-band count rates (0.5--8.0 keV) within the aperture of our sources and converted these values to absorbed X-ray flux $F_{\rm X, abs}$ upper limits using WebPIMMS\footnote{http://heasarc.nasa.gov/Tools/w3pimms.html}, assuming the emission is from a $Z = 0.2 Z_{\sun}$ metallicity plasma with $kT_{\rm X}=0.15$ keV. We then corrected for absorption to derive unabsorbed (emitted) X-ray fluxes $F_{\rm X, unabs}$, assuming an absorbing column equal to the weighted average $N_{\rm H}$ in the source direction, given by the \cite{lab05} survey of Galactic neutral hydrogen. Finally, we simulated spectra of the $Z = 0.2 Z_{\sun}$, $kT_{\rm X}=0.15$ keV plasma to determine the emission measure $EM_{\rm X}$ (and consequently, the electron density $n_{\rm X} = \sqrt{EM_{\rm X}/V}$). The results of these analyses for the 15 SMC \hii\ regions are listed in Table~\ref{table:PXupperlimits}.

\subsection{Errors Associated with Each Term} \label{sec:uncertainty}

Each pressure term calculated using the methods described above will have an associated error, and there are many uncertainties which will contribute given the variety of data and analyses required. Nonetheless, we attempt to assess these errors in the following ways. For the direct radiation pressure $P_{\rm dir}$, the dominant uncertainty is the relation of $L_{\rm H\alpha}$ to $L_{\rm bol}$, as described in Section~\ref{sec:Pdir}. Thus, for our error bars on $P_{\rm dir}$ have incorporated the factor of 2 uncertainty in the conversion of $L_{\rm H\alpha}$ to $L_{\rm bol}$. Our calculation of $P_{\rm IR}$ is fairly robust, and the largest error comes from the 2\% uncertainty in the {\it Spitzer} photometry, which corresponds to a 2.8\% error in the flux ratios of Figure~\ref{fig:LMC_models}. Therefore, we interpolated the $U$--$q_{\rm PAH}$ grid for $\pm$2.8\% of our flux ratios to obtain a corresponding error in $U$. These uncertainties lead to errors of the order 5--10\% in $P_{\rm IR}$.

In the case of $P_{\rm HII}$, we have uncertainty in the flux density $F_{\nu}$ over the radii of our \hii\ regions due to the low resolution of the radio data. Therefore, we have measured $F_{\nu}$ for $\pm$one resolution element in our radio image and obtained the corresponding uncertainty in $n_{\rm e}$. This error is relatively small, $\sim$10--15\% in $n_{\rm e}$ and $P_{\rm HII}$. Finally, the range of $P_{\rm X}$ is given by the uncertainty in the X-ray spectral fits of emission measure (and correspondingly, the hot gas density $n_{\rm X}$) and of the temperature $kT_{\rm X}$. We employ these 90\% confidence limits derived in our spectral fits, as listed in Table~\ref{table:PXresults}. Generally, the density $n_{\rm X}$ was poorly constrained in lower signal sources (e.g., N4, N30, and N59), as further evidenced by the poor reduced chi-squared values in those fits. Therefore, in some cases, the error bars on $P_{\rm X}$ can be relatively large, although the typical uncertainties were around $\sim$30--50\% in $n_{\rm X}$. 

\section{Results} \label{sec:results}

\begin{figure}
\begin{center}
\includegraphics[width=\columnwidth]{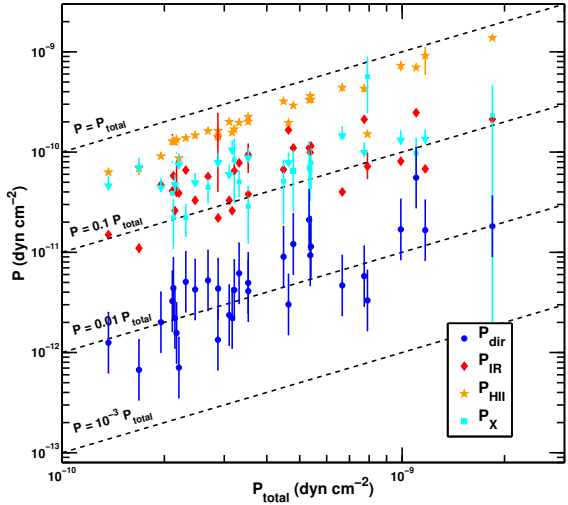}
\end{center}
\caption{Individual pressure terms and associated uncertainties versus the total pressure $P_{\rm tot}$ for the 32 \hii\ regions. Dashed lines are meant to show how much each term contributes to the total pressure. The light blue arrows represent the $P_{\rm X}$ upper limits of the 15 SMC \hii\ regions that are not detected in archival {\it XMM-Newton} and {\it Chandra} data; for our calculation of $P_{\rm total}$, we assume the SMC $P_{\rm X}$ upper limits are the pressures of the hot gas. Section~\ref{sec:uncertainty} describes how error bars were calculated for each term.}
\label{fig:PdirvsP}
\end{figure}

\begin{figure}
\begin{center}
\includegraphics[width=\columnwidth]{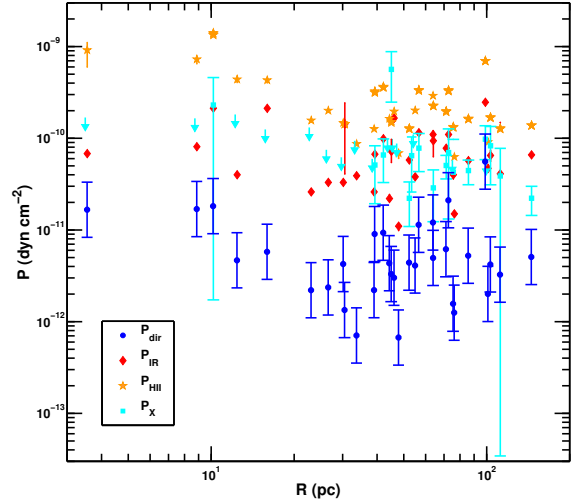}
\end{center}
\caption{Pressures versus HII region size $R$ of the 32 \hii\ regions. The light blue arrows represent the $P_{\rm X}$ upper limits of the 15 SMC \hii\ regions that are not detected in archival {\it XMM-Newton} and {\it Chandra} data. See Section~\ref{sec:uncertainty} for how error bars were assessed for each term.} 
\label{fig:PvsR}
\end{figure}

Following the multi-wavelength analyses performed above, we calculate the pressure associated with the direct stellar radiation pressure $P_{\rm dir}$, the dust-processed radiation pressure $P_{\rm IR}$, the warm ionized gas pressure $P_{\rm HII}$, and the hot X-ray gas pressure $P_{\rm X}$. Table~\ref{tab:Presults} gives the pressure components and associated errors measured for all the \hii\ regions, and Figure~\ref{fig:PdirvsP} plots the pressure terms versus their sum, $P_{\rm total}$, to facilitate visual comparison of the parameters. As shown in Figure~\ref{fig:PvsR}, we do not find any trends in the pressure terms versus size $R$ of the \hii\ regions. In all the targets except one, $P_{\rm HII}$ dominates over $P_{\rm IR}$ and $P_{\rm X}$. The exception is N191, which has a $P_{\rm IR}$ roughly equal to its $P_{\rm HII}$, although the errors on $P_{\rm IR}$ are quite large. For all sources detected in the X-rays except N30, $P_{\rm HII}$ is a factor 2--7 above $P_{\rm X}$ and $P_{\rm IR} \gs P_{\rm X}$ in all sources. Broadly, the relation between the terms is $P_{\rm HII} > P_{\rm IR} > P_{\rm X} > P_{\rm dir}$. In the entire sample, $P_{\rm dir}$ is 1--2 orders of magnitude smaller than the other pressure components. We note that while $P_{\rm dir} > P_{\rm HII}$ at distances $\ls$75 pc from R136 in the giant \hii\ region 30 Doradus \citep{lopez11}, the warm ionized gas is what is driving the expansion currently and dominates the energetics when averaged over the entire source. 

\begin{deluxetable*}{lcccc} 
\tablecolumns{5}
\tablewidth{0pt} \tablecaption{Pressure Results\tablenotemark{a}} 
\tablehead{\colhead{Source} & \colhead{$P_{\rm dir}$} & \colhead{$P_{\rm IR}$} & \colhead{$P_{\rm HII}$} & \colhead{$P_{\rm X}$} \\
\colhead{} &  \colhead{($\times 10^{-12}$ dyn cm$^{-2}$)} & \colhead{($\times 10^{-10}$ dyn cm$^{-2}$)} & \colhead{($\times 10^{-10}$ dyn cm$^{-2}$)} & \colhead{($\times 10^{-10}$ dyn cm$^{-2}$)}
}
\startdata
\cutinhead{LMC Sources} 
N4 & 18.2$^{+18.2}_{-9.1}$ & 2.13$^{+0.08}_{-0.07}$ & 13.8$\pm0.1$ & 2.31$\pm$2.29 \\
N11 & 5.08$^{+5.08}_{-2.54}$ & 0.66$^{+0.03}_{-0.01}$ & 1.38$\pm0.01$ & 0.22$\pm$0.08 \\
N30 & 3.31$^{+3.31}_{-1.65}$ & 0.72$^{+0.26}_{-0.18}$ & 1.51$^{+0.04}_{-0.03}$ & 5.64$\pm$3.17 \\
N44 & 4.21$^{+4.21}_{-2.10}$ & 0.65$\pm0.09$ & 1.69$^{+0.01}_{-0.02}$ & 0.83$\pm$0.52 \\
N48 & 1.57$^{+1.57}_{-0.78}$ & 0.40$\pm0.04$ & 1.33$^{+0.01}_{-0.02}$ & 0.43$\pm$0.43 \\
N55 & 4.41$^{+4.41}_{-2.20}$ & 0.58$^{+0.03}_{-0.02}$ & 1.28$^{+0.01}_{-0.02}$ & 0.22$\pm$0.11 \\
N59 & 11.4$^{+11.4}_{-5.70}$ & 1.15$^{+0.03}_{-0.04}$ & 3.35$^{+0.02}_{-0.04}$ & 0.78$\pm$0.35 \\
N79 & 4.96$^{+4.96}_{-2.48}$ & 0.94$^{+0.26}_{-0.31}$ & 2.25$^{+0.03}_{-0.01}$ & 0.29$\pm$0.16 \\
N105 & 9.34$^{+9.34}_{-4.67}$ & 0.99$\pm0.04$ & 3.63$^{+0.02}_{-0.06}$ & 0.66$\pm$0.33 \\
N119 & 5.24$^{+5.24}_{-2.62}$ & 0.57$^{+0.01}_{-0.02}$ & 1.62$^{+0.02}_{-0.01}$ & 0.44$\pm$0.13 \\
N144 & 6.18$^{+6.18}_{-3.09}$ & 0.78$\pm0.03$ & 1.97$^{+0.01}_{-0.03}$ & 0.51$\pm$0.14 \\
30 Dor & 55.7$^{+55.7}_{-27.8}$ & 2.47$^{+0.08}_{-0.09}$ & 6.99$^{+0.02}_{-0.04}$ & 0.98$\pm$0.39 \\
N160 & 21.1$^{+21.1}_{-10.5}$ & 1.10$^{+0.04}_{-0.05}$ & 3.32$^{+0.03}_{-0.05}$ & 0.70$\pm$0.57 \\
N180 & 9.03$^{+9.03}_{+4.52}$ & 0.67$\pm0.03$ & 3.21$^{+0.04}_{-0.06}$ & 0.51$\pm$0.32 \\
N191 & 1.34$^{+1.34}_{-0.67}$ & 1.43$^{+1.00}_{-1.02}$ & 1.43$\pm0.01$ & -- \\
N206 & 3.26$^{+3.26}_{-1.63}$ & 0.41$^{+1.08}_{-0.40}$ & 1.28$^{+0.01}_{-0.02}$ & 0.39$\pm$0.39 \\
\cutinhead{SMC Sources} 
DEM~S74 & 0.67$^{+0.67}_{-0.34}$ & 0.11$\pm0.01$ & 0.69$^{+0.04}_{-0.09}$ & $<$0.88 \\
N13 & 16.9$^{+16.9}_{-8.5}$ & 0.81$^{+0.04}_{-0.03}$ & 7.28$^{+0.59}_{-0.78}$ & $<$1.65 \\
N17 & 2.37$^{+2.37}_{-1.18}$ & 0.33$^{+0.02}_{-0.01}$ & 2.00$^{+0.06}_{-0.07}$ & $<$0.75 \\
N19 & 4.67$^{+4.67}_{-2.34}$ & 0.40$^{+0.03}_{-0.01}$ & 4.40$^{+0.37}_{-0.34}$ & $<$1.82 \\
N22 & 5.78$^{+5.78}_{-2.89}$ & 2.12$^{+0.12}_{-0.04}$ & 4.31$^{+0.24}_{-0.29}$ & $<$1.25 \\
N36 & 4.34$^{+4.34}_{-2.17}$ & 0.22$^{+0.02}_{-0.01}$ & 1.63$^{+0.04}_{-0.03}$ & $<$0.99 \\
N50 & 1.25$^{+1.25}_{-0.63}$ & 0.15$\pm0.01$ & 0.63$\pm0.01$ & $<$0.58 \\
N51 & 0.71$^{+0.71}_{-0.35}$ & 0.39$\pm0.02$ & 0.87$\pm0.01$ & $<$0.94 \\
N63 & 2.20$^{+2.20}_{-1.10}$ & 0.26$^{+0.01}_{-0.02}$ & 1.57$^{+0.05}_{-0.06}$ & $<$1.31 \\
N66 & 12.1$^{+12.1}_{-6.04}$ & 1.10$^{+0.06}_{-0.04}$ & 2.92$\pm0.04$ & 0.65$\pm$0.39 \\
N71 & 16.6$^{+16.6}_{-8.32}$ & 0.68$\pm0.03$ & 9.16$^{+1.90}_{-3.18}$ & $<$1.69 \\
N76 & 4.10$^{+4.10}_{-2.05}$ & 0.38$\pm0.02$ & 2.01$^{+0.03}_{-0.04}$ & $<$1.10 \\
N78 & 3.02$^{+3.02}_{-1.51}$ & 1.66$^{+0.09}_{-0.05}$ & 1.96$\pm0.03$ & $<$0.98 \\
N80 & 2.21$^{+2.21}_{-1.11}$ & 0.26$^{+0.02}_{-0.01}$ & 1.27$\pm0.02$ & $<$0.60 \\
N84 & 2.01$^{+2.01}_{-1.00}$ & 0.47$\pm0.02$ & 0.91$\pm0.01$ & $<$0.55 \\
N90 & 4.25$^{+4.25}_{+2.13}$ & 0.33$\pm0.02$ & 1.47$\pm0.08$ & $<$0.62 \\
\enddata
\tablenotetext{a}{See Section~\ref{sec:uncertainty} for how error bars were assessed for each term.} 
\label{tab:Presults}
\end{deluxetable*}

\section{Discussion} \label{sec:discussion}

\subsection{The Importance of Direct Radiation Pressure} \label{sec:radpressure}

From Section~\ref{sec:results}, it is evident that direct radiation pressure does not play a significant role in the dynamics of the regions. However, given the age and size of our sources, they are too large/evolved for the radiation pressure to be significant. The reason is that the pressure terms have a different radial dependence: $P_{\rm dir} \propto r_{\rm HII}^{-2}$, while $P_{\rm HII} \propto r_{\rm HII}^{-3/2}$, where $r_{\rm HII}$ is the shell radius. One can obtain a rough estimate of the characteristic radius $r_{\rm ch}$ where a given source transitions from radiation-pressure driven to gas-pressure driven by setting the total radiation pressure (i.e., the direct radiation as well as the dust-processed radiation) equal to the warm gas pressure and solving for $r_{\rm ch}$. In this case, we find 

\begin{equation}
r_{\rm ch} = \frac{\alpha_{\rm B}}{12 \pi \phi} \bigg( \frac{\epsilon_{\rm 0}}{2.2 k_{\rm B} T_{\rm HII}} \bigg)^2 f_{\rm trap,tot}^{2} \frac{\psi^2 S}{c^{2}},
\label{eq:rch}
\end{equation} 

\noindent
where $\epsilon_{\rm 0} = 13.6$ eV, the photon energy necessary to ionize hydrogen, $\alpha_{\rm B}$ is the case-B recombination coefficient, and $\phi$ is a dimensionless quantity which accounts for dust absorption of ionizing photons and for free electrons from elements besides hydrogen. In a gas-pressure dominated \hii\ region, $\phi$ = 0.73 if He is singly ionized and 27\% of photons are absorbed by dust \citep{mw97}. The $f_{\rm trap,tot}$ represents the factor by which radiation pressure is enhanced by trapping energy in the shell through several mechanisms, including trapping of stellar winds, infrared photons, and Ly$\alpha$ photons. Here, we adopt $f_{\rm trap,tot} = 2$, as in \cite{km09}, although we note this factor is uncertain and debated, as discussed in Section~\ref{sec:dusty}. Lastly, $\psi$ is the ratio of bolometric power to the ionizing power in a cluster; we set $\psi = 3.2$ using the $\langle S \rangle/ \langle M_{*} \rangle$ and the $\langle L \rangle/ \langle M_{*} \rangle$ relations of \cite{mr10}. Using these values, the above equation reduces to 

\begin{equation} \label{eq:rch_S49}
r_{\rm ch} = 0.072~S_{49} ~~{\rm pc},
\end{equation}

\noindent
where $S$ is the ionizing photon rate, and $S_{49} \equiv S/10^{49}$ s$^{-1}$. We note that the derivation of Equations~\ref{eq:rch} and~\ref{eq:rch_S49} required several simplifying assumptions (e.g., regarding the coupling of the radiation to dust), and thus the estimate of $r_{\rm ch}$ should be viewed as a rough approximation of the true radius when an \hii\ region transitions from radiation- to gas-pressure dominated. 

We can estimate $S_{49}$ for our \hii\ regions based on their H$\alpha$ luminosity \citep{mw97}: 

\begin{equation}
L_{\rm H\alpha} = 1.04 \times 10^{37} S_{49} ~~{\rm erg~s}^{-1}. 
\label{eq:S}
\end{equation}

\noindent
We list the resulting ionizing photon rates $S$ for our sample in Table~\ref{tab:extinction}. Given these values, we find a range $r_{\rm ch} \sim$ 0.01--7 pc for 31 \hii\ regions and $r_{\rm ch} \approx$ 33 pc for 30 Dor. As our sample have radii $\sim$10--150 pc, the 32 \hii\ regions are much too large to be radiation-pressure dominated at this stage. 

This result demonstrates the need to investigate young, small \hii\ regions to probe radiation pressure dominated sources. The best candidates would be hypercompact (HC) \hii\ regions, which are characterized by their very small radii $\ls$0.05 pc and high electron densities $n_{\rm e} \gs 10^{6}$ cm$^{-3}$ \citep{hoare07}. HC \hii\ regions may represent the earliest evolutionary phase of massive stars when they first begin to emit Lyman continuum radiation, and thus they offer the means to explore the dynamics before the thermal pressure of the ionized gas dominates. 

Giant \hii\ regions which are powered by more massive star clusters may also be radiation pressure dominated. For example, \cite{km09} showed that the super star clusters (with masses $M \sim 10^{5}-10^{6} M_{\sun}$) in the starburst galaxy M82 are likely radiation pressure dominated.

\subsection{Hot Gas Leakage from HII Shells} \label{sec:leakage}

In Section~\ref{sec:results}, we have demonstrated that the average X-ray gas pressure $P_{\rm X}$ is below the $10^{4}$ K gas pressure $P_{\rm HII}$. For the X-ray detected \hii\ regions, the median $P_{\rm X}/P_{\rm HII}$ is 0.22, with a range in $P_{\rm X}/P_{\rm HII} \sim$ 0.13--0.50 (excluding N30, which has $P_{\rm X}/P_{\rm HII} \approx 3.7\pm2.1$).  For the 15 non-detected sources, we set upper limits on $P_{\rm X}$ requiring at least 13 of the 15 \hii\ regions to have  $P_{\rm X} / P_{\rm HII} < 1$ and nine to have $P_{\rm HII} \gs 2 P_{\rm X}$. 

The low $P_{\rm X}$ values we derive are likely due to the partial/incomplete confinement of the hot gas by the \hii\ shells (e.g., \citealt{rosen14}). If completely confined by an \hii\ shell expanding into a uniform density ISM, the hot gas pressure $P_{\rm X}$ would be large \citep{castor75,weaver77}. Conversely, a freely expanding wind would produce a negligible $P_{\rm X}$ \citep{chev85}. In the intermediate case, a wind bubble expands into an inhomogeneous ISM, creating holes in the shell where the hot gas can escape and generating a moderate $P_{\rm X}$. For example, \cite{harperclark09} argue the Carina nebula is experiencing hot gas leakage based partly on its observed X-ray gas pressure of $P_{\rm X} \sim2 \times 10^{-10}$ dyn cm$^{-2}$, whereas the complete confinement model predicts $P_{\rm X} \sim 10^{-9}$ dyn cm$^{-2}$ and the freely expanding wind model predicts $P_{\rm X} \sim 10^{-13}$ dyn cm$^{-2}$ for Carina. 

Recent observational and theoretical evidence has emerged that hot gas leakage may be a common phenomenon. Simulations have demonstrated that hot gas leakage can be significant through low-density pores in molecular material \citep{tenorio07,dale08,rogers13}. Observationally, signatures of hot gas leakage in individual \hii\ regions has been noted based on their X-ray luminosities and morphologies, such as in M17 and the Rosette Nebula \citep{townsley03}, the Carina Nebula \citep{harperclark09}, and 30 Dor \citep{lopez11}. The results we have presented here on a large sample demonstrate that hot gas leakage may be typical among evolved \hii\ regions, implying that the mechanical energy injected by winds and SNe can be lost easily without doing work on the shells. 

\subsection{How Much Momentum Can Be Imparted to Gas by Dust-Processed Radiation?} \label{sec:dusty}

Although we have found that the warm gas pressure $P_{\rm HII}$ dominates at the shells of our sources, a couple \hii\ regions (N191 in the LMC and N78 in the SMC, although we caution that the uncertainty in $P_{\rm IR}$ in N191 is large) have nearly comparable $P_{\rm IR}$ and $P_{\rm HII}$, and all 32 sources have $P_{\rm IR} \gg P_{\rm dir}$. Physically, this scenario can occur if the shell is optically thick to the dust-processed IR photons, amplifying the exerted force of those photons. In all 32 regions of our sample, the amplification factor caused by trapping the photons $f_{\rm trap, IR} \equiv P_{\rm IR}/P_{\rm dir}$ is quite large, with $f_{\rm trap, IR} \sim$ 4--100 and a median value of $f_{\rm trap, IR} \sim$ 10. 

From a theoretical perspective, it has been debated in the literature how much momentum can be deposited in matter by IR photons. \cite{km09} argued that the imparted momentum would be limited to $f_{\rm trap,IR} \ls$ a few because holes in the shell would cause the radiation to leak out of those pores. Conversely, if every photon is absorbed many times, then all the energy of the radiation field is converted to kinetic energy of the gas; this scenario imparts the most momentum to the shell. An intermediate case is in optically thick systems, where photons are absorbed at least once, and the momentum deposition is dependent on the optical depth $\tau_{\rm IR}$ of the region \citep{thompson05,murray10,andrews11}. 

Recent simulations by \cite{krumholz12b,krumholz13} indicate that $f_{\rm trap, IR}$ can be large as long as the radiation flux is below a critical value that depends on the dust optical depth. This critical value corresponds to the radiation flux being large enough so that the pressure of the dust-trapped radiation field is at the same order of magnitude as the gas pressure. At fluxes above the critical value, a radiation-driven Rayleigh-Taylor (RRT) instability develops and severely limits the value of $f_{\rm trap, IR}$ by creating low-density channels through which radiation can escape. For example, in one case in \cite{krumholz12} where the RRT instability does not develop, they obtain $f_{\rm trap, IR} \approx 90$, whereas when the radiation flux is increased so that radiation forces  become significant and there is instability, $f_{\rm trap, IR}$ drops to a few. Clearly in the case of our sources, we are in the regime where the radiation pressure is not dominant compared to the warm gas pressure, and RRT instability is not expected (though two of our sources are near the threshold of instability). Thus, the high values of $f_{\rm trap, IR}$ we obtain are consistent with these models. 

\section{Summary} \label{sec:summary}

In this paper, we have performed a systematic, multi wavelength analysis of 32 \hii\ regions in the Magellanic Clouds to assess the role of stellar feedback in their dynamics. We have employed optical, IR, radio, and X-ray images to measure the pressures associated with direct stellar radiation, dust-processed radiation, warm ionized gas, and hot X-ray emitting plasma at the shells of these sources. We have found that the warm ionized gas dominates over the other terms in all sources, although two \hii\ regions have comparable dust-processed components. The hot gas pressures are relatively weaker, and the direct radiation pressures are 1--2 orders of magnitude below the other terms. 

We explore three implications to this work. First, we emphasize that younger, smaller \hii\ regions, such as hypercompact \hii\ regions, should be studied to probe the role of direct radiation pressure and the hot gas at early times. Secondly, the low X-ray luminosities and pressures we derive indicate the hot gas is only partially confined in all of our sources, suggesting that hot gas leakage is a common phenomenon in evolved \hii\ regions. Finally, we have demonstrated that the dust-processed component can be significant and comparable to warm gas pressure, even if the direct radiation pressure is comparatively less. These observational results are consistent with recent numerical work showing that the dust-processed component can be largely amplified as long as it does not drive winds.

\acknowledgements

Support for this work was provided by National Aeronautics and Space Administration through Chandra Award Number GO2--13003A and through Smithsonian Astrophysical Observatory contract SV3--73016 to MIT and UCSC issued by the Chandra X-ray Observatory Center, which is operated by the Smithsonian Astrophysical Observatory for and on behalf of NASA under contract NAS8--03060. Support for LAL was provided by NASA through the Einstein Fellowship Program, grant PF1--120085, and the MIT Pappalardo Fellowship in Physics. MRK acknowledges the Alfred P. Sloan Foundation, NSF CAREER grant AST--0955300, and NASA ATP grant NNX13AB84G. ADB acknowledges partial support from a Research Corporation for Science Advancement Cottrell Scholar Award and the NSF CAREER grant AST--0955836. ERR acknowledges support from the David and Lucile Packard Foundation and NSF grant AST--0847563. DC acknowledges support for this work provided by NASA through the Smithsonian Astrophysical Observatory contract SV3--73016 to MIT for support of the Chandra X-ray Center, which is operated by the Smithsonian Astrophysical Observatory for and on behalf of NASA under contract NAS8--03060. 

\begin{appendix}

\section{The Filling Factor of the Hot Gas} \label{app:hot gas}

The conversion of emission measure $EM_{\rm X}$ to hot gas electron density $n_{\rm X}$ requires an assumption about the volume occupied by the hot gas, parametrized by a filling factor $f_{\rm X}$. For a fixed gas temperature $kT_{\rm X}$ (which is determined from the spectral fitting and is independent of the assumed $f_{\rm X}$), the inferred density and pressure scale as $f_{\rm X}^{-1/2}$. One can attempt to deduce $f_{\rm X}$ from a combination of morphology and spectral modeling (as in e.g., \citealt{pellegrini11}). However, for the purposes of understanding the global dynamics, this approach can be misleading, as we demonstrate here. Following the reasoning outlined below, we set $f_{\rm X} = 1$.

We are interested in the global dynamics of the regions, which are described by the virial theorem. Neglecting magnetic fields (which may not be negligible, but we lack an easy means to measure them), the Eulerian form of the virial theorem is \citep{mckee92}:

\begin{equation}
\frac{1}{2}\ddot{I} = 2(\mathcal{T} - \mathcal{T}_s) + \mathcal{R} - \mathcal{R}_s + \mathcal{W} - \frac{1}{2} \frac{d}{dt} \int_S (\rho\mathbf{v} r^2) \cdot d\mathbf{S}, 
\end{equation}

\noindent
where

\begin{eqnarray}
I & = & \int_V \rho r^2 \, dV \\
\mathcal{T} & = & \frac{1}{2} \int (3P + \rho v^2) \, dV \\
\mathcal{T}_s & = & \frac{1}{2} \int_S \mathbf{r}\cdot\mathbf{\Pi}\cdot d\mathbf{S} \\
\mathcal{R} & = & \int_V u_{\rm rad} \, dV \\
\mathcal{R}_s & = & \int_S \mathcal{r}\cdot \mathbf{P}_{\rm rad} \cdot d\mathbf{S} \\
\mathcal{W} & = & -\int \rho \mathbf{r}\cdot\nabla\phi \, dV. 
\end{eqnarray}

Here, $V$ is the volume, $S$ is the surface of this volume, $\rho$, $\mathbf{v}$, and $P$ are the gas density, velocity, and pressure, $\mathbf{\Pi} = \rho\mathbf{vv} + P\mathbf{I}$ is the fluid pressure tensor, $u_{\rm rad}$ is the frequency-integrated radiation energy density, $\mathbf{P}_{\rm rad}$ is the radiation pressure tensor, $\phi$ is the gravitational potential, and $\mathbf{I}$ is the identity tensor. The terms $I$, $\mathcal{T}$, $\mathcal{R}$, and $\mathcal{W}$ may be identified, respectively, as the moment of inertia, the total thermal plus kinetic energy, the total radiation energy, and the gravitational binding energy. The terms subscripted with $s$ represent external forces exerted at the surface of the volume, and are likely negligible in comparison with the internal terms for an H~\textsc{ii} region with large energy input by massive stars.

Since manifestly $\ddot{I}$ either is very positive now, or was in the recent past (otherwise the shell would not have expanded), the goal of this work is to understand the balance between the various positive terms on the right-hand side of the equation. The terms $P_{\rm IR}$ and $P_{\rm dir}$ are simply two different parts of $\mathcal{R}$, corresponding to energy stored in different parts of the electromagnetic spectrum, while $P_{\rm HII}$ and $P_{\rm X}$ are part of $\mathcal{T}$. Writing out the virial theorem in this manner makes the importance of the filling factor clear. The term we are interested in evaluating is the kinetic plus thermal energy of the X-ray emitting gas,

\begin{equation}
\mathcal{T}_X = \frac{3}{2} \int P_X \, dV = \langle P_X \rangle V,
\end{equation}

\noindent
where we have dropped the $\rho v^2$ term on the assumption that the flow velocity is subsonic with respect to the hot gas sound speed, and in the second step we have defined the volume-averaged pressure $\langle P_{\rm X} \rangle$, as distinct from the local pressure at a given point. The quantity $\langle P_{\rm X} \rangle$ can be understood as the partial pressure of the hot gas, including proper averaging down for whatever volumes it does not occupy. Thus we see that the quantity of interest is \textit{not} the local number density or pressure of the hot gas, it is the volume-averaged or partial pressure. Now recall that, for fixed $T_{\rm X}$ and fixed observed emission measure, local pressure scales with filling factor as $P_{\rm X} \propto f_{\rm X}^{-1/2}$, so a small volume filling factor increases $P_{\rm X}$. However, since the volume occupied by the hot gas scales as $f_X$, it follows that $\mathcal{T}_{\rm X} \propto \langle P_{\rm X} \rangle \propto f_{\rm X}^{1/2}$, i.e., a small volume filling factor implies that the hot gas is less, not more, important for the large-scale dynamics.

This analysis has two important implications. First, the choice that makes the hot gas as dynamically-important as possible is to set $f_X = 1$, i.e., to assume that the hot gas fills most of the available volume. In this case we simply have $P_{\rm X} = \langle P_{\rm X} \rangle$, and this is the choice we make in this work. A detailed assessment of $f_{\rm X}$ that gives a value $\ll 1$, as performed by \citet{pellegrini11}, can imply an even smaller dynamical role for the hot gas, but not a larger one (although understanding of filling factors is important for other considerations, such as the internal dynamics of \hii\ regions). The second implication is that it is inconsistent to treat $P_{\rm X}$ as the quantity of interest for the global dynamics while simultaneously adopting $f_{\rm X} < 1$. Once can certainly attempt to measure $f_X$ and thus obtain a more accurate assessment of $P_{\rm X}$, but in this case the quantities that should be compared with other pressures is $\langle P_{\rm X} \rangle = f_{\rm X} P_{\rm X}$, \textit{not} $P_{\rm X}$. The volume-averaged pressure is the relevant quantity for global dynamics, not the local pressure. We note that the above discussion of the filling factor applies to the warm gas as well, and we have also assumed a filling factor of order unity for the warm 10$^{4}$ K gas. 

\end{appendix}

\end{document}